\documentclass[opre,nonblindrev]{informs3} 

\OneAndAHalfSpacedXI 

\usepackage{endnotes}
\let\footnote=\endnote

\usepackage{natbib}
\usepackage{multirow,float,ctable}
 \bibpunct[, ]{(}{)}{,}{a}{}{,}%
 \def\bibfont{\small}%

\TheoremsNumberedThrough     
\ECRepeatTheorems

\EquationsNumberedThrough    

\begin{document}

\RUNAUTHOR{Zavala, Kim, Anitescu, and Birge}

\MANUSCRIPTNO{}

\RUNTITLE{Stochastic Market Clearing with Consistent Pricing}

\TITLE{A Stochastic Electricity Market Clearing Formulation with Consistent Pricing Properties\thanks{Preprint ANL/MCS-P5110-0314}}

\ARTICLEAUTHORS{%
\AUTHOR{\bf Victor M. Zavala}
\AFF{Department of Chemical and Biological Engineering, University of Wisconsin-Madison\\ 1415 Engineering Dr, Madison, WI 53706 \EMAIL{\{victor.zavala@wisc.edu\}}} 
\AUTHOR{\bf Kibaek Kim, Mihai Anitescu}
\AFF{Mathematics and Computer Science Division, Argonne National Laboratory\\ 9700 South Cass Avenue, Argonne, IL 60439 \EMAIL{\{kibaekkim@mcs.anl.gov,anitescu@mcs.anl.gov\}}} 
\AUTHOR{\bf John Birge}
\AFF{The University of Chicago Booth School of Business\\
5807 South Woodlawn Avenue, Chicago, IL 60637 \EMAIL{jbirge@chicagobooth.edu}}
} 

\ABSTRACT{%
We argue that deterministic market clearing formulations introduce arbitrary distortions between day-ahead and expected real-time prices that bias economic incentives and block diversification. We extend and analyze the stochastic clearing formulation proposed by \citet{philpott} in which the social surplus function induces penalties between day-ahead and real-time quantities. We prove that the formulation yields price distortions that are bounded by the bid prices, and we show that adding a similar penalty term to transmission flows and phase angles ensures boundedness throughout the network. We prove that when the price distortions are zero, day-ahead quantities converge to the quantile of real-time counterparts. The undesired effects of price distortions suggest that stochastic settings provide significant benefits over deterministic ones that go beyond social surplus improvements. We propose additional metrics to evaluate these benefits.  
}

\HISTORY{This paper was first submitted on May, 2014; last revised on October 25, 2015.}

\maketitle

\section{Introduction}

Day-ahead markets enable commitment and pricing of resources to hedge against uncertainty in demand, generation, and network capacities that are observed in real time. The day-ahead market is cleared by independent system operators (ISOs) using deterministic unit commitment (UC) formulations that rely on expected capacity forecasts, while uncertainty is handled by allocating reserves that are used to balance the system if real-time capacities deviate from the forecasts. A large number of deterministic clearing formulations have been proposed in the literature. Representative examples include those of \citet{carrion,gribikramp}, and \citet{ucbook}. Pricing issues arising in deterministic clearing formulations have been explored by \citet{gribikelmp,galianasocial}, and \citet{oneill}.  

In addition to guaranteeing reliability and maximizing social surplus, several metrics are monitored by ISOs to ensure that the market operates efficiently. For instance, as is discussed in \citet{ott}, the ISO must ensure that market players receive economic incentives that promote participation (give participants the incentive to follow commitment and dispatch signals). It is also desired that day-ahead and real-time prices are sufficiently close or converge. One of the reasons is that price convergence is an indication that capacity forecasts are effective reflections of real-time capacities. Recent evidence provided by \citet{bowden1}, however, has shown that persistent and predictable deviations between day-ahead and real-time prices (premia) exist in certain markets.  This can bias the incentives to a subset of players and block the entry of new players and technologies.  The introduction of purely financial players was intended to eliminate premia, but recent evidence provided by \citet{birgevirtual} shows that this has not been fully effective. One hypothesis is that virtual players can exploit predictable price differences in the day-ahead market to create artificial congestion and benefit from financial transmission rights \citep{joskow2000transmission}. 

Prices are also monitored by ISOs to ensure that they do not run into financial deficit (a situation called revenue inadequacy) when balancing payments to suppliers and from consumers. This is discussed in detail in \citet{philpottftr}. In addition, ISOs might need to use uplift payments and adjust prices to protect suppliers from operating at an economic loss. This is necessary to prevent players from leaving the market. As discussed by \citet{oneill}, \cite{morales2012pricing}, and \citet{gribikelmp}; uplift payments can result from using incomplete characterizations of the system in the clearing model. Such characterizations can arise, for instance, in the presence of nonconvexities and stochasticity. 

Achieving efficient market operations under intermittent renewable generation is a challenge for the ISOs because uncertainty follows complex spatiotemporal patterns not faced before (\citet{zavalauc}). In addition, the power grid is relying more strongly on natural gas and transportation infrastructures, and it is thus necessary to quantify and mitigate uncertainty in more systematic ways (\citet{liugas,zavalagas}). 

\subsection{Previous Work}

A wide range of stochastic formulations of day-ahead market clearing and operational UC procedures has been previously proposed. In operational UC models, on/off decisions are made in advance (here-and-now) to ensure that enough running capacity is available at future times to balance the system. The objective of these formulations is to ensure reliability and maximization of social surplus (cost in case of inelastic demands) in intra-day operations. Examples include the works of \citet{takritibirge,wang,zavalauc,ryan,ruiz,bouffardstochastic,papavasiliou2013multiarea}. These studies have demonstrated significant improvements in reliability over deterministic formulations. However, these works do not explore pricing issues.                                                                                                                                                                                                                                                                                                    

Stochastic day-ahead clearing formulations have been proposed by \citet{kaye} and \citet{fullerpricing}.  \citet{kaye} analyse day-ahead and real-time markets under uncertainty and argue that day-ahead prices should be set to expected values of the real-time prices. This {\em price consistency} ensures that the day-ahead market does not bias real-time market incentives in the long run. Such consistency also avoids arbitrage as is argued by \citet{oren2013}.\citet{fullerpricing} propose pricing schemes to achieve cost recovery for all suppliers (i.e., payments cover the suppliers production costs). The pricing schemes, however, rely only partially on dual variables generated by the stochastic clearing model which are adjusted to achieve cost recovery. Consequently, these procedures do not guarantee dual and model consistency. 

\citet{morales2012pricing} propose a stochastic clearing model to price electricity in pools with stochastic producers. Their model co-optimizes energy and reserves and they prove that it leads to revenue adequacy in expectation. In addition, they prove that prices allow for cost recovery in expectation for all players (i.e., no uplifts are needed in expectation) but pricing consistency is not explored. 

\citet{philpott} propose a stochastic formulation that captures day-ahead and real-time bidding of both suppliers and consumers. The formulation maximizes the day-ahead social surplus and the expected value of the real-time corrections by considering the possibility of players' bidding in the real-time market. The authors prove that the formulation leads to revenue adequacy in expectation and provide conditions under which adequacy will hold for each scenario. The authors do not explore pricing consistency and economic incentives. 

\citet{oren2013} propose a stochastic equilibrium formulation in which players bid parameters of a quadratic supply function to maximize the expected value of their profit function while the ISO uses these parameters to solve the clearing model and generate day-ahead and real-time quantities and prices. It is shown that the equilibrium model generates day-ahead prices that converge to expected value prices and thus achieve consistency. It is also shown that day-ahead quantities converge to expected value quantities and a small case study is presented to demonstrate that the formulation yields higher social surplus and producer profits compared to deterministic clearing. The proposed formulation uses a quadratic supply function and quadratic penalties for deviations between day-ahead and real-time quantities. No network and no capacity constraints are considered. 

\citet{morales2013pricing} propose a bilevel stochastic optimization formulation that uses forecast capacities of stochastic suppliers as degrees of freedom. Using computational studies, they demonstrate that their framework provides cost recovery for all suppliers and for each scenario. The authors, however, do not discuss the effects of the modified pricing strategy on consumer payments (the demands are treated as inelastic) and no theoretical guarantees are provided. In particular, it is not guaranteed that a set of day-ahead capacities and prices exist that can achieve cost recovery for both suppliers and consumers in each scenario. While plausible, we believe that this requires further evidence and theoretical justification.  

\subsection{Contributions of This Work}

In this work, we argue that deterministic formulations generate day-ahead prices that are distorted representations of expected real-time prices. This pricing inconsistency arises because solving a day-head clearing model using summarizing statistics of uncertain capacities (e.g., expected forecasts) does not lead to day-ahead prices that are expected values of the real-time prices. We argue that these price distortions lead to diverse issues such as the need of uplift payments as well as arbitrary and biased incentives that block diversification. We extend and analyze the stochastic clearing formulation of \citet{philpott} in which linear supply functions for day-ahead and real-time markets are used. The structure of this surplus function has the key property that yields bounded price distortions. We also prove that when the price distortion is zero, the formulation yields day-ahead quantities that converge to the quantile of their real-time counterparts. In addition, we prove that the formulation yields revenue adequacy in expectation and yields zero uplifts in expectation. We provide several case studies to demonstrate the properties of the stochastic formulation. 

The paper is structured as follows. In Section \ref{sec:setting} we describe the market setting. In Section \ref{sec:models} we present deterministic and stochastic formulations of the day-ahead ISO clearing problem.  In Section \ref{sec:metrics} we present a set of performance metrics to assess the benefits of the stochastic formulation over its deterministic counterpart. In Section \ref{sec:properties} we present the pricing properties of the formulation. In Section \ref{sec:computation} we present case studies to demonstrate the developments. Concluding remarks and directions of future work are provided in Section \ref{sec:conclusions}. 
\section{Market Setting}\label{sec:setting}

We consider a market setting based on the work of \citet{philpott} and \citet{ott}. A set of suppliers (generators) $\mathcal{G}$ and consumers (demands) $\mathcal{D}$ bid into the day-ahead market by providing price bids $\alpha_i^{g}\geq 0$, $i\in\mathcal{G}$ and $\alpha_j^{d}\geq 0$, $j\in\mathcal{D}$, respectively. If a given demand is inelastic, we set the bid price to $\alpha_{j}^{d}=VOLL$ where $VOLL$ denotes the value of lost load, typically 1,000 \$/MWh. Suppliers and consumers also provide estimates of the available capacities $\bar{g}_{i}$ and $\bar{d}_{j}$, respectively. We assume that these capacities satisfy $0\leq \bar{g}_i\leq Cap_i^g$ and $0\leq \bar{d}_j\leq Cap_j^d$ where $0\leq Cap_{i}^g < +\infty$ is the total installed capacity of the supplier (its maximum possible supply) and $0\leq Cap_{i}^d < +\infty$ is the total installed capacity of the consumer (its maximum possible demand). The cleared day-ahead quantities for suppliers and consumers are given by $g_{i}$ and $d_{j}$, respectively. These satisfy $0\leq g_{i}\leq \bar{g}_{i}$ and $0\leq d_{j}\leq \bar{d}_{j}$. 

Suppliers and consumers are connected through a network
comprising of a set of lines $\mathcal{L}$ and a set
of nodes $\mathcal{N}$.  For each line $\ell \in\mathcal{L}$
we define its sending node as $snd(\ell)\in\mathcal{N}$ and
its receiving node as $rec(\ell)\in\mathcal{N}$ (we highlight that this definition of sending node is arbitrary because the flow can go in both directions). For each
node $n\in\mathcal{N}$, we define its set of receiving lines
as $\mathcal{L}_n^{rec}\subseteq \mathcal{L}$ and its set of
sending lines as $\mathcal{L}_{n}^{snd}\subseteq \mathcal{L}$. These sets are given by
\begin{subequations}
\begin{align}
  \mathcal{L}_n^{rec}&=\{\ell\in\mathcal{L}\,|\,n=rec(\ell)\},\; n\in\mathcal{N}\\
  \mathcal{L}_n^{snd}&=\{\ell\in\mathcal{L}\,|\,n=snd(\ell)\},\; n\in\mathcal{N}.
\end{align}
\end{subequations}
Day-ahead capacities $\bar{f}_{\ell}$ are also typically estimated for the transmission lines. We assume that these satisfy $0\leq \bar{f}_{\ell}\leq Cap^f_{\ell}$. Here, $0\leq Cap^f_\ell< +\infty$ is the installed capacity of line (its maximum possible value). The cleared day-ahead flows are given by $f_\ell$ such that $-\bar{f}_\ell\leq f_\ell\leq \bar{f}_\ell$. 
The flows $f_\ell$ are determined by the line susceptance $B_\ell$ and the phase angle difference between two nodes of the line. Day-ahead capacities $\underline\theta_n,\bar\theta_n$ are estimated for each node $n\in\mathcal{N}$. The cleared day-ahead phase angles are given by $\theta_n$ such that $\underline\theta_n \leq \theta_n \leq \bar\theta_n$ for $n\in\mathcal{N}$. We define the set of all suppliers connected to node $n\in\mathcal{N}$ as $\mathcal{G}_n\subseteq \mathcal{G}$ and the set of demands connected to node $n$ as $\mathcal{D}_n\subseteq \mathcal{D}$. Subindex $n(i)$ indicates the node at which supplier $i\in \mathcal{G}$ is connected, and $n(j)$ indicates the node at which the demand $j\in\mathcal{D}$ is connected. We use subindex $i$ exclusively for suppliers and subindex $j$ exclusively for consumers. 

At the moment the day-ahead market is cleared, the
real-time market conditions are uncertain. In particular, we
assume that a subset of generation, demand, and transmission
line capacities are uncertain. We further assume that discrete 
distributions comprising a finite set of scenarios
$\Omega$ and $p(\omega)$ denote the probability of
scenario $\omega\in\Omega$. We also require that
$\sum_{\omega\in\Omega}p(\omega)=1$. The expected value of
variable $Y(\cdot)$ is given by
$\mathbb{E}[Y(\omega)]=\sum_{\omega \in \Omega}
p(\omega)Y(\omega)$. If $Y(\omega)$ is scalar-valued, the quantile function $\mathbb{Q}$ is defined as
\begin{align}\label{eq:defQuantile}
  \mathbb{Q}_{Y(\omega)}(p) := \inf \left\{ y \in \mathbb{R} \;:\; \mathbb{P}(Y(\omega) \leq y) \geq p\right\}.
\end{align}
Moreover, the median is denoted as $\mathbb{M}[Y(\omega)] = \mathbb{Q}_{Y(\omega)}(0.5)$ and satisfies
\begin{align}
\mathbb{M}[Y(\omega)]&= \mathop{\textrm{argmin}}_{m} \mathbb{E}[|Y(\omega)-m|],\label{eq:median} 
\end{align}
where $|\cdot|$ is the absolute value function.

In the real-time market, the suppliers can offer to sell additional generation over the agreed day-ahead quantities at a bid price $\alpha_i^{g,+}\geq 0$. The additional generation is given by $(G_{i}(\omega)-g_{i})_+$ where $G_{i}(\omega)$ is the cleared quantity in the real-time market and $0\leq \bar{G}_{i}(\omega)\leq Cap^g_i$ is the realized capacity under scenario $\omega\in{\Omega}$. Real-time generation quantities are bounded as $0\leq G_{i}(\omega) \leq \bar{G}_{i}(\omega)$. Here, $(X-x)_+:=\textrm{max}\{X-x,0\}$. The suppliers also have the option of buying electricity at an offering price $\alpha_i^{g,-}\geq 0$ to account for any uncovered generation ($G_{i}(\omega)-g_{i})_-$ over the agreed day-ahead quantities. Here, $(X-x)_-=\textrm{max}\{-(X-x),0\}$.  

Consumers provide bid prices $\alpha_j^{d,-}\geq 0$ to buy additional demand $(D_{j}(\omega)-d_{j})_+$ in the real-time market, where $D_{j}(\omega)$ is the cleared quantity and $0\leq \bar{D}_{j}(\omega)\leq Cap^d_{j}$ is the available demand capacity realized under scenario $\omega\in\Omega$. We thus have $0\leq D_{j}(\omega)\leq \bar{D}_{j}(\omega)$. Consumers also have the option of selling the demand deficit $(D_{j}(\omega)-d_{j})_-$ at price $\alpha_{j}^{d,+}\geq 0$. 

The flows cleared in the real-time market are given by $F_{\ell}(\omega)$ and satisfy $-\bar{F}_{\ell}(\omega)\leq {F}_{\ell}(\omega)\leq \bar{F}_{\ell}(\omega)$. Here, $\bar{F}_{\ell}(\omega)$ is the transmission line capacity realized under scenario $\omega\in\Omega$ and satisfies $-Cap^f_{\ell}\leq \bar{F}_{\ell}(\omega)\leq Cap^f_{\ell}$. Uncertain line capacities can be used to model $N-x$ contingencies or uncertainties in capacity due to ambient conditions (e.g., ambient temperature affects line capacity).   The cleared phase angles in the real-time market are given by $\Theta_n(\omega)$ such that $\underline\theta_n\leq \Theta_n(\omega) \leq \bar\theta_n$ for $n\in\mathcal{N}$.

We also define day-ahead clearing prices (i.e., locational marginal prices) for each node $n\in\mathcal{N}$ as $\pi_n$. The real-time prices are defined as $\Pi_n(\omega), \; \omega\in \Omega$. 
\section{Clearing Formulations}\label{sec:models}

In this section, we present {\em energy-only} day-ahead deterministic and stochastic clearing formulations. The term ``energy-only'' indicates that no unit commitment decisions are made. We consider these simplified formulations in order to focus on important concepts related to pricing and payments to suppliers and consumers. Model extensions are left as a topic of future research.  

\subsection{Deterministic Formulation}

In a deterministic setting, the day-ahead market is cleared by solving the following optimization problem.
\begin{subequations}\label{eq:detday-ahead}
\begin{align}
  \min_{d_j,g_i,f_{\ell},\theta_n} \quad
  & \sum_{i \in \mathcal{G}} \alpha_{i}^gg_{i}-\sum_{j \in \mathcal{D}}\alpha_j^dd_{j}\\
  \text{s.t.} \quad
  & \sum_{\ell \in \mathcal{L}_n^{rec}} f_{\ell} -\sum_{\ell \in \mathcal{L}_n^{snd}} f_{\ell} + \sum_{i \in \mathcal{G}_n} g_{i} - \sum_{i \in \mathcal{D}_n} d_{i}=0,\quad (\pi_n)\quad n\in\mathcal{N}\label{eq:detdaflow}\\
  & f_\ell = B_\ell (\theta_{rec(\ell)} - \theta_{snd(\ell)}), \quad \ell\in\mathcal{L} \label{eq:powerflow}\\
  & -\bar{f}_{\ell}\leq f_{\ell} \leq \bar{f}_{\ell}, \quad \ell \in \mathcal{L} \label{eq:det:flowbounds}\\
  & 0\leq g_{i}\leq \bar{g}_{i},\quad i\in\mathcal{G}\\
  & 0\leq d_{j}\leq \bar{d}_{j},\quad j\in\mathcal{D}\\
  & \underline\theta_n \leq \theta_n \leq \overline\theta_n, \quad n\in\mathcal{N}.\label{eq:det:anglebounds}
\end{align}
\end{subequations}
The objective function of this problem is the day-ahead {\em negative} social surplus. The solution of this problem gives the day-ahead quantities $g_{i},d_{j}$, flows $f_{\ell}$, phase angles $\theta_n$, and prices $\pi_n$. The deterministic formulation assumes a given value for the capacities $\bar{g}_{i},\bar{d}_{j},\bar{f}_{\ell}, \underline\theta_n$, and $\overline\theta_n$. Because the conditions of the real-time market are uncertain at the time the day-ahead problem \eqref{eq:detday-ahead} is solved, these capacities are typically assumed to be the most probable ones (e.g., the expected value or {\em forecast} for supply and demand capacities) or are set based on the current state of the system (e.g., for line capacities and phase angle ranges). In particular, it is usually assumed that $\bar{g}_{i}=\mathbb{E}[\bar{G}_{i}(\omega)]$, $\bar{d}_{j}=\mathbb{E}[\bar{D}_{j}(\omega)]$, and $\bar{f}_{\ell}$ is the most probable state.  One can also assume that $\bar{g}_{i}=Cap^g_i$ and $\bar{d}_j=Cap^d_j$, and $\bar{f}_{\ell}=Cap^f_{\ell}$. Such an assumption, however, can yield high economic penalties if the day-ahead dispatched quantities are far from those realized in the real-time market. Similarly, one can also assume conservative capacities (e.g., worst-case). In this sense, the day-ahead capacities $\bar{g}_i,\bar{d}_j,\bar{f}_{\ell}$ can be used as mechanisms to hedge against risk, as experienced ISO operators do to allow for a safety margin. Doing so, however, gives only limited control because the players need to summarize the entire possible range of real-time capacities in one statistic. In Section \ref{sec:metrics} we argue that this limitation can induce a distortion between day-ahead and real-time prices and biases revenues. 

When the capacities become known, the ISO uses fixed day-ahead committed quantities $g_{i},d_{j},f_{\ell},\theta_n$, to solve the following real-time clearing problem.
\begin{subequations}\label{eq:det_recourse}
\begin{align}
  \min_{D_j(\cdot),G_i(\cdot),F_{\ell}(\cdot),\Theta_n(\cdot)} \quad 
  & \sum_{i \in \mathcal{G}} \left(\alpha_{i}^{g,+}(G_{i}(\omega)-g_{i})_+ - \alpha_{i}^{g,-}(G_{i}(\omega)-g_{i})_-\right)\\
  & \qquad +\sum_{j \in \mathcal{D}} \left(\alpha_{j}^{d,+}(D_{j}(\omega)-d_{j})_--\alpha_{j}^{d,-}(D_{j}(\omega)-d_{j})_+ \right)\\
  \text{s.t.} \quad
  & \sum_{\ell \in \mathcal{L}_n^{rec}} F_{\ell}(\omega) -\sum_{\ell \in \mathcal{L}_n^{snd}} F_{\ell}(\omega) + \sum_{i \in \mathcal{G}_n} G_{i}(\omega) - \sum_{j \in \mathcal{D}_n} D_{j}(\omega)=0,\quad (\Pi_n(\omega)), \quad n\in\mathcal{N}\label{eq:detrtflow}\\
  & F_\ell(\omega) = B_\ell (\Theta_{rec(\ell)}(\omega) - \Theta_{snd(\ell)}(\omega)), \quad n\in\mathcal{N}\\
  & -\bar{F}_{\ell}(\omega)\leq F_{\ell}(\omega) \leq \bar{F}_{\ell}(\omega), \quad \ell \in \mathcal{L}\\
  & 0\leq G_{i}(\omega)\leq \bar{G}_{i}(\omega),\quad i\in\mathcal{G}\\
  & 0\leq D_{j}(\omega)\leq \bar{D}_{j}(\omega),\quad j\in\mathcal{D}\\
  & \underline\theta_n \leq \Theta_n(\omega) \leq \overline\theta_n, \quad n\in\mathcal{N}.
\end{align}
\end{subequations}
The objective function of this problem is the real-time negative social surplus. The solution of this problem yields different real-time quantities ${G}_{i}(\omega),{D}_{j}(\omega)$, flows $F_{\ell}(\omega)$, phase angles $\Theta_n(\omega)$, and prices $\Pi_n(\omega)$ depending on the scenario $\omega\in\Omega$ realized. 

\subsection{Stochastic Formulation}

Motivated by the structure of the day-ahead and real-time market problems, we consider the stochastic market clearing formulation:
\begin{subequations}\label{eq:stoch}
\begin{align}
  \min_{d_j,D_j(\cdot),g_i,G_i(\cdot),f_{\ell},F_{\ell}(\cdot),\theta_n,\Theta_n(\cdot)} \;
  & \varphi^{sto} := \sum_{i \in \mathcal{G}} \mathbb{E}\left[\alpha_{i}^gg_{i}+\alpha_{i}^{g,+}(G_{i}(\omega)-g_{i})_+ - \alpha_{i}^{g,-}(G_{i}(\omega)-g_{i})_-\right]\nonumber\\
  &\qquad + \sum_{j \in \mathcal{D}} \mathbb{E}\left[-\alpha_j^dd_{j}+\alpha_{j}^{d,+}(D_{j}(\omega)-d_{j})_- - \alpha_{j}^{d,-}(D_{j}(\omega)-d_{j})_+\right]\nonumber\\
  & \qquad + \sum_{\ell\in\mathcal{L}} \mathbb{E}\left[\Delta\alpha^{f,+}_{\ell} (F_{\ell}(\omega)-f_{\ell})_+ + \Delta\alpha^{f,-}_{\ell} (F_{\ell}(\omega)-f_{\ell})_-\right] \notag\\
  & \qquad + \sum_{n\in\mathcal{N}} \mathbb{E}\left[\Delta\alpha^{\theta,+}_n (\Theta_n(\omega) - \theta_n)_+ + \Delta\alpha^{\theta,-}_n (\Theta_n(\omega) - \theta_n)_-\right]\\
  \text{s.t.} \;
  & \sum_{\ell \in \mathcal{L}_n^{rec}} f_{\ell} -\sum_{\ell \in \mathcal{L}_n^{snd}} f_{\ell} + \sum_{i \in \mathcal{G}_n} g_{i} - \sum_{i \in \mathcal{D}_n} d_{i}=0,\quad(\pi_n)\quad n\in\mathcal{N}\label{eq:networkfwd}\\
  & f_\ell = B_\ell (\theta_{rec(\ell)} - \theta_{snd(\ell)}), \quad \ell\in\mathcal{L} \label{eq:DAangle}\\
  & \sum_{\ell \in \mathcal{L}_n^{rec}} \left(F_{\ell}(\omega)-f_{\ell}\right) -\sum_{\ell \in \mathcal{L}_n^{snd}} \left(F_{\ell}(\omega)-f_{\ell}\right)+ \sum_{i \in \mathcal{G}_n} \left(G_{i}(\omega)-g_{i}\right) \nonumber\\
  & \qquad - \sum_{j \in \mathcal{D}_n} \left(D_{j}(\omega)-d_{j}\right)=0,\quad(p(\omega)\Pi_n(\omega))\quad \omega \in \Omega,n\in\mathcal{N}\label{eq:networkrt}\\
  & F_\ell(\omega) = B_\ell (\Theta_{rec(\ell)}(\omega) - \Theta_{snd(\ell)}(\omega)), \quad \omega\in\Omega,\ell\in\mathcal{L}\label{eq:RTangle}\\
  & -\bar{F}_{\ell}(\omega)\leq F_{\ell}(\omega) \leq \bar{F}_{\ell}(\omega), \;\; \omega\in\Omega,\ell \in \mathcal{L}\label{eq:boundstoch1}\\
  & 0\leq G_{i}(\omega)\leq \bar{G}_{i}(\omega),\qquad \omega\in\Omega,i\in\mathcal{G}\\
  & 0\leq D_{j}(\omega)\leq \bar{D}_{j}(\omega),\qquad \omega\in\Omega,j\in\mathcal{D}\label{eq:boundstoch2}\\
  & \underline\theta_n \leq \Theta_n(\omega) \leq \overline\theta_n, \quad \omega\in\Omega,n\in\mathcal{N}.
\end{align}
\end{subequations}

The stochastic setting provides a natural mechanism to {\em anticipate} the effects of day-ahead decisions on real-time market corrections. This property gives rise to several important pricing and payment properties, as we will see in the following section.   

The above formulation is partially based on the one proposed by \citet{philpott}. We highlight the following features of the model:
\begin{itemize}

\item The real-time prices (duals of the network balance \eqref{eq:networkrt}) have been weighted by their corresponding probabilities. This feature will enable us to construct the Lagrange function of the problem in terms of expectations. 

\item The network balance in the real-time market is written in terms of the residual quantities $(G_{i}(\omega)-g_{i})$, $(D_{j}(\omega)-d_{j})$, and flows $(F_{\ell}(\omega)-f_{\ell})$. This feature will be key in obtaining consistent prices and it emphasizes the fact that the real-time market is a market of corrections.

\item We assume that the real-time quantity bounds $\bar{G}_{i}(\omega),\bar{D}_{j}(\omega),\bar{F}_{\ell}(\omega)$ are independent of the day-ahead quantities.

\end{itemize}

The differences between the proposed formulation and the one presented by \citet{philpott} are the following.
\begin{itemize}
\item The formulation does not impose bounds on the day-ahead quantities, flows and phase angles. In Section \ref{sec:properties} we will prove that the penalization terms render bounds for the day-ahead quantities, flows and phase angles redundant (see Theorem~\ref{th:networkbounds}).

\item  The parameters $\Delta\alpha^{f,+}_{\ell},\Delta\alpha^{f,-}_{\ell}, \Delta\alpha^{\theta,+}, \Delta\alpha^{\theta,-}> 0$ penalize deviations between day-ahead and real-time quantities. In Section \ref{sec:metrics} we will see that these penalties are motivated by the structure of the social surplus and in Section \ref{sec:properties}  we will show that they are key to ensure desirable pricing properties.

\item We allow for randomness in the transmission line capacities. In Section \ref{sec:properties} we will see that doing so has no effect on the underlying properties of the model. 

\item We assume that the stochastic problem has relative complete recourse. That is, there exist a feasible real-time recourse decision for any day-ahead decision.

\end{itemize}

We refer to the solution of the stochastic formulation \eqref{eq:stoch} as the {\em here-and-now} solution to reflect the fact that a single implementable decision must be made now in anticipation of the uncertain future and that day-ahead quantities and flows are scenario-independent. We also consider the (ideal, non-implementable) {\em wait-and-see} (WS) solution. For details, refer to \citet{birge}. In the WS setting, we assume that the capacities for each scenario are actually known at the moment of decision. In other words, we assume availability of perfect information. In order to obtain the WS solution, the clearing problem \eqref{eq:stoch} is solved by allowing first-stage decisions $g_i,d_j,f_{\ell}$ to be scenario-dependent. It is not difficult to prove that in this case, each scenario generates day-ahead prices and quantities that are equal to real-time counterparts because no corrections are necessary. We denote the expected social surplus obtained under perfect information as $\varphi^{sto}_{WS}$.  
\section{ISO Performance Metrics for Market Clearing} \label{sec:metrics}

In this section, we discuss some objectives of the ISOs from a market operations standpoint and use these to motivate a new set of metrics to quantify performance of deterministic and stochastic formulations. We place special emphasis on the structure of the {\em social surplus function} and on the issue of {\em price consistency}. We provide arguments as to why price consistency is a key property in achieving incentives. We argue that deterministic formulations do not actually yield price consistency and hence result in a range of undesired effects such as biased payments, revenue inadequacy, and the need for uplifts. 

We highlight that we define different metrics based on market behavior in expectation. A practical way of interpreting these {\em expected metrics} is the following: assume that the market conditions of a given day are repeated over a sequence of days and we collect the results over such period by using each day as a scenario. We then compute a certain metric (like the social welfare) to perform the comparisons between the stochastic and deterministic clearing mechanisms to evaluate performance. In this sense, market behavior in expectation can also interpreted as long run market behavior. 

\subsection{Social Surplus}

Consider the combination of the day-ahead and real-time costs for suppliers and consumers,
\begin{subequations}\label{costs}
\begin{align}
C^g_{i}(\omega)&=+\alpha_{i}^gg_{i}+\alpha_{i}^{g,+}(G_{i}(\omega)-g_{i})_+ - \alpha_{i}^{g,-}(G_{i}(\omega)-g_{i})_-\\
C^d_{j}(\omega)&=-\alpha_j^dd_{j}+\alpha_{j}^{d,+}(D_{j}(\omega)-d_{j})_- - \alpha_{j}^{d,-}(D_{j}(\omega)-d_{j})_+.
\end{align}
\end{subequations}

We define the {\em incremental bid prices} as $\Delta\alpha_i^{g,+} := \alpha_i^{g,+} - \alpha_i^g$, $\Delta\alpha_i^{g,-} := \alpha_i^{g} - \alpha_i^{g,-}$, $\Delta\alpha_j^{d,+} := \alpha_j^{d,+} - \alpha_j^d$ and $\Delta\alpha_j^{d,-} := \alpha_j^{d} - \alpha_j^{d,-}$. To avoid degeneracy, we require that the incremental bid prices are positive: $\Delta\alpha_i^{g,+}, \Delta\alpha_i^{g,-}, \Delta\alpha_j^{d,+}, \Delta\alpha_j^{d,-} > 0$

\begin{theorem}
Assume that the incremental bid prices are positive. The cost functions for suppliers and consumers can be expressed as
\begin{subequations}\label{eq:asymcost}
\begin{align}
  C_{i}^g(\omega) &= \alpha_i^{g} G_{i}(\omega) + \Delta \alpha_{i}^{g,+} (G_{i}(\omega)-g_{i})_+ + \Delta \alpha_{i}^{g,-} (G_{i}(\omega)-g_{i})_-, \quad i\in\mathcal{G},\omega\in\Omega\\
  C_{j}^d(\omega) &= -\alpha_{j}^{d} D_{j}(\omega) + \Delta \alpha_{j}^{d,+} (D_{j}(\omega)-d_{j})_- + \Delta \alpha_{j}^{d,-} (D_{j}(\omega)-d_{j})_+, \quad j\in\mathcal{D},\omega\in\Omega.
\end{align} 
\end{subequations}
\end{theorem}
\proof{Proof}
Consider the cost function for suppliers
\begin{align*}
  C^g_{i}(\omega) &= \alpha_{i}^g g_{i} + \alpha_{i}^{g,+} (G_{i}(\omega)-g_{i})_+ - \alpha_{i}^{g,-} (G_{i}(\omega)-g_{i})_-\\
  &= \alpha_{i}^g g_{i} + (\alpha_{i}^{g} + \Delta \alpha_i^{g,+}) (G_{i}(\omega)-g_{i})_+ - (\alpha_{i}^{g} - \Delta \alpha_i^{g,-}) (G_{i}(\omega)-g_{i})_- \\
  &= \alpha_{i}^g g_{i} + \alpha_{i}^{g} (G_{i}(\omega) - g_{i})_+ - \alpha_{i}^{g} (G_{i}(\omega) - g_{i})_- + \Delta \alpha_i^{g,+} (G_{i}(\omega) - g_{i})_+ + \Delta \alpha_i^{g,-} (G_{i}(\omega) - g_{i})_- \\
  &= \alpha_{i}^g g_{i} + \alpha_{i}^{g} (G_{i}(\omega) - g_{i}) + \Delta \alpha_i^{g,+} (G_{i}(\omega) - g_{i})_+ + \Delta \alpha_i^{g,-} (G_{i}(\omega) - g_{i})_- \\
  &= \alpha_{i}^{g} G_{i}(\omega) + \Delta \alpha_i^{g,+} (G_{i}(\omega) - g_{i})_+ + \Delta \alpha_i^{g,-} (G_{i}(\omega) - g_{i})_-.
\end{align*}
The last two equalities follow from the fact that $X-x=(X-x)_+-(X-x)_-$. The same property applies to $C^d_j(\omega)$ (using the appropriate cost terms).\Halmos
\endproof

We say that the incremental bid prices are {\em symmetric} if $\Delta\alpha_i^{g,+}=\Delta\alpha_i^{g,-}$ and $\Delta\alpha_j^{d,+}=\Delta\alpha_j^{d,-}$. Denote the symmetric prices by $\Delta\alpha_i^g := \Delta\alpha_i^{g,+}=\Delta\alpha_i^{g,-}$ and $\Delta\alpha_j^d := \Delta\alpha_j^{d,+}=\Delta\alpha_j^{d,-}$.
\begin{corollary}\label{thm:abscost}
If the incremental bid prices are symmetric, then the cost functions for suppliers and consumers can be expressed as
\begin{subequations}\label{eq:abscost}
\begin{align}
  C_{i}^g(\omega) &= \alpha_i^{g} G_{i}(\omega) + \Delta \alpha_{i}^{g} |G_{i}(\omega)-g_{i}|, \quad i\in\mathcal{G},\omega\in\Omega\\
  C_{j}^d(\omega) &= -\alpha_{j}^{d} D_{j}(\omega) + \Delta \alpha_{j}^{d} |D_{j}(\omega)-d_{j}|, \quad j\in\mathcal{D},\omega\in\Omega.
\end{align} 
\end{subequations}
\end{corollary}
\proof{Proof}
Consider the cost function for suppliers
\begin{align*}
  C^g_{i}(\omega) &= \alpha_{i}^{g} G_{i}(\omega) + \Delta \alpha_i^{g,} (G_{i}(\omega) - g_{i})_+ + \Delta \alpha_i^{g,} (G_{i}(\omega) - g_{i})_- \\
  &= \alpha_i^{g} G_{i}(\omega) + \Delta \alpha_{i}^{g} |G_{i}(\omega)-g_{i}|,
\end{align*}
because of the fact that $|X-x|=(X-x)_++(X-x)_-$. The same property applies to $C_j^d(\omega)$ (using the appropriate cost terms).\Halmos
\endproof

\begin{definition}[Social Surplus]
We define the {\em expected negative social surplus} (or {\em social surplus} for short) as
\begin{align}\label{eq:surplus}
  \varphi &:= \mathbb{E}\left[\sum_{i \in \mathcal{G}} C^g_{i}(\omega)+\sum_{j \in \mathcal{D}}C_{j}^d(\omega)\right]\nonumber\\
  &=\varphi^g+\varphi^d,
\end{align}
where $\varphi^g,\varphi^d$ are the {\em expected supply and consumer costs},
\begin{subequations}\label{eq:expectedcost}
\begin{align}
  \varphi^g &:= \mathbb{E}\left[\sum_{i \in \mathcal{G}} C^g_{i}(\omega)\right] \notag\\
  &= \sum_{i\in\mathcal{G}} \left( \alpha_i^{g} \mathbb{E}[G_i(\omega)] + \Delta\alpha_i^{g,+} \mathbb{E}[(G_{i}(\omega) - g_i)_+] + \Delta\alpha_i^{g,-} \mathbb{E}[(G_{i}(\omega) - g_i)_-] \right) \\
  \varphi^d &:= \mathbb{E}\left[\sum_{j \in \mathcal{D}} C^d_{j}(\omega)\right] \notag\\
  &= \sum_{j\in\mathcal{D}} \left( -\alpha_{j}^{d} \mathbb{E}[D_j(\omega)] + \Delta\alpha_j^{d,+} \mathbb{E}\left[ (D_{j}(\omega) - d_j)_-\right] + \Delta\alpha_j^{d,-} \mathbb{E}\left[ (D_{j}(\omega) - d_j)_+\right] \right).
\end{align}
\end{subequations}

\end{definition}

This particular structure of the expected social surplus function was noticed by \citet{philpott} and provides interesting insights. From Equation \eqref{eq:expectedcost}, we note that the expected quantities $\mathbb{E}[G_i(\omega)]$, $\mathbb{E}[D_j(\omega)]$ act as forecasts of the day-ahead quantities and are priced by using the day-ahead bids $\alpha_i^g,\alpha_d^j$ (first term). This immediately suggests that it is the {\em expected cleared quantities} $G_i(\omega),D_j(\omega)$ and {\em not} the capacities $\bar{g}_i,\bar{d}_j$ that are to be used as forecasts, as is done in the day-ahead deterministic formulation \eqref{eq:detday-ahead}. The second and third terms penalize deviations of the real-time quantities from the day-ahead commitments using the incremental bid prices. More interestingly, Corollary~\ref{thm:abscost} suggests that when the incremental bid prices are symmetric (i.e., $\Delta\alpha_i^{g,+}=\Delta\alpha_i^{g,-}$ and $\Delta\alpha_j^{d,+}=\Delta\alpha_j^{d,-}$), day-ahead quantities will tend to converge to the median of the real-time quantities if the expected social surplus function is minimized. A deterministic setting, however, cannot guarantee optimality in this sense because it minimizes the day-ahead and real-time components of the surplus function {\em separately}. In particular, the expected social surplus for the deterministic formulation is obtained by solving the day-ahead problem \eqref{eq:detday-ahead} followed by the solution of the real-time problem \eqref{eq:det_recourse} for all scenarios $\omega \in \Omega$. The day-ahead surplus and the expected value of the real-time surplus are then combined to obtain the expected surplus $\varphi$. 

A deterministic setting can yield surplus inefficiencies because it cannot properly anticipate the effect of day-ahead decision on real-time market decisions. For instance, certain suppliers can be inflexible in the sense that they cannot modify their day-ahead supply easily in the real-time market (e.g., coal plants). This results in constraints of the form $g_i=G_i(\omega)$ or $d_j=D_j(\omega),\omega \in \Omega$. This inflexibility can trigger inefficiencies because the operator is forced to use expensive units in the real-time market (e.g., combined-cycle) or because load shedding is needed to prevent infeasibilities. Most studies on stochastic market clearing and unit commitment have focused on showing improvements in social surplus over deterministic formulations. In Section \ref{sec:computation} we demonstrate that even when social surplus differences are negligible, the resulting prices and payments can be drastically different. This situation motivates us to consider alternative metrics for monitoring performance. 

We note that the objective function of the stochastic clearing formulation \eqref{eq:stoch} can be written as
\begin{align}
  \varphi^{sto} = \varphi 
  &+ \sum_{\ell\in\mathcal{L}} \mathbb{E}\left[\Delta\alpha^{f,+}_{\ell} (F_{\ell}(\omega)-f_{\ell})_+ + \Delta\alpha^{f,-}_{\ell} (F_{\ell}(\omega)-f_{\ell})_-\right] \notag\\
  &+ \sum_{n\in\mathcal{N}} \mathbb{E}\left[\Delta\alpha^{\theta,+}_n (\Theta_n(\omega) - \theta_n)_+ + \Delta\alpha^{\theta,-}_n (\Theta_n(\omega) - \theta_n)_-\right],\label{eq:surplussto}
\end{align}
where $\varphi$ is the expected negative surplus function defined in \eqref{eq:surplus}.  Consequently, if $\Delta\alpha^{f,+}_{\ell},\Delta\alpha^{f,-}_{\ell}, \Delta\alpha^{\theta,+}, \Delta\alpha^{\theta,-}$ are sufficiently small, we have that $\varphi^{sto}\approx \varphi$. 

\subsection{Pricing Consistency}\label{sec:priceconsistency}

We seek that the day-ahead prices be consistent representations of the expected real-time prices. In other words, we seek that the {\em expected price distortions} (also known as {\em expected price premia}) $\pi_n-\mathbb{E}[\Pi_n(\omega)],\; n\in\mathcal{N}$ be zero or at least in a bounded neighborhood. This is desired for various reasons that we will explain. 

\begin{definition}[Price Distortions] We define the {\em expected price distortion} or {\em expected price premia} as 
\begin{align}
\mathcal{M}^{\pi}_n:=\pi_n-\mathbb{E}\left[\Pi_n(\omega)\right], \quad n\in\mathcal{N}.
\end{align}
We say that the {\em price is consistent} at node $n\in\mathcal{N}$ if $\mathcal{M}^{\pi}_n=0$. In addition, we define the {\em node average} and {\em maximum absolute distortions}, 
\begin{subequations}
\begin{align}
\mathcal{M}^{\pi}_{avg}&:=\frac{1}{|\mathcal{N}|}\sum_{n\in\mathcal{N}}|\mathcal{M}^{\pi}_n|\\
\mathcal{M}^{\pi}_{max}&:=\mathop{\textrm{max}}_{n\in\mathcal{N}}|\mathcal{M}^{\pi}_n|.
\end{align}
\end{subequations}
\end{definition}

Pricing consistency is related to the desire that day-ahead and real-time prices converge, as is discussed by \citet{ott}. Note, however, that it is unrealistic to expect that day-ahead and real-time prices converge in each scenario. This is possible only in the absence of uncertainty (capacity forecasts are perfect such as in the perfect information setting). Any real-time deviation in capacity from a day-ahead forecast will lead to a deviation between day-ahead and real-time prices. It is possible, however, to ensure that {\em day-ahead and real-time prices converge in expectation}. This situation also implies that any deviation of the real-time price from the day-ahead price is entirely the result of {\em unpredictable random factors}. This is also equivalent to saying that day-ahead prices converge to the expected value of the real-time prices.  

Pricing consistency cannot be guaranteed with deterministic formulations because the day-ahead clearing model forecasts real-time capacities, not real-time quantities. Consequently, players are forced to ``summarize'' their possible real-time capacities in single statistics $\bar{d}_j,\bar{g}_i,\bar{f}_{\ell}$. Expected values are typically used. This summarization, however, is inconsistent because it does not effectively average real-time market performance as the structure of the surplus function \eqref{eq:expectedcost} suggests. In fact, as we show in Section \ref{sec:properties}, expected values need not be the right statistic to use in the day-ahead market.  This is consistent with the observations made by \citet{morales2013pricing}. In addition, we note that certain random variables might be difficult to summarize (e.g., if they follow multimodal and heavy-tailed distributions). For instance, consider that there is uncertainty about the state of a transmission line in the real-time market (i.e., there is a probability that it will fail). In a deterministic setting it is difficult to come up with a "forecast" value for the day-ahead capacity $\bar{f}_\ell$ in such a case. 

\subsection{Suppliers and Consumer Payments}

As argued by \citet{kaye}, we can justify the desire of seeking price consistency by analyzing the {\em payments} to the market players. The payment includes the day-ahead settlement plus the correction payment given at real-time prices, as is the standard practice in market operations. For more details, see \citet{ott} and \citet{philpott}. 

\begin{definition}[Payments] The payments to suppliers and from consumers in scenario $\omega \in \Omega$ are defined as follows:
\begin{subequations}
\begin{align}
P_{i}^g(\omega)&:=g_{i}\pi_{n(i)}+(G_{i}(\omega)-g_{i})\Pi_{n(i)}(\omega)\nonumber\\
&=g_{i}(\pi_{n(i)}-\Pi_{n(i)}(\omega))+G_{i}(\omega)\Pi_{n(i)}(\omega),\quad i\in\mathcal{G},\omega\in \Omega\\
P_{j}^d(\omega)&:=-d_{j}\pi_{n(i)}-(D_{j}(\omega)-d_{j})\Pi_{n(j)}(\omega)\nonumber\\
&=d_{j}(\Pi_{n(i)}(\omega)-\pi_{n(i)})-D_{j}(\omega)\Pi_{n(j)}(\omega),\quad j\in\mathcal{D},\omega\in \Omega.
\end{align}
\end{subequations}
We say that the {\em expected payments are consistent} if they satisfy
\begin{subequations}\label{eq:payconsistency}
\begin{align}
\mathbb{E}\left[P_{i}^g(\omega)\right]&=+\mathbb{E}\left[G_{i}(\omega)\Pi_{n(i)}(\omega)\right], \quad i\in\mathcal{G}\\
\mathbb{E}\left[P_{j}^d(\omega)\right]&=-\mathbb{E}\left[D_{j}(\omega)\Pi_{n(j)}(\omega)\right], \quad j\in\mathcal{D},
\end{align}
\end{subequations}
where
\begin{subequations}\label{eq:expectedpayments}
\begin{align}
\mathbb{E}\left[P_{i}^g(\omega)\right]&=+g_{i}\mathcal{M}_{n(i)}^{\pi}+\mathbb{E}\left[G_{i}(\omega)\Pi_{n(i)}(\omega)\right], \quad i\in\mathcal{G}\\
\mathbb{E}\left[P_{j}^d(\omega)\right]&=-d_{j}\mathcal{M}_{n(j)}^{\pi}-\mathbb{E}\left[D_{j}(\omega)\Pi_{n(j)}(\omega)\right], \quad j\in\mathcal{D}.
\end{align}
\end{subequations}
\end{definition}

If the prices are consistent at each node $n\in\mathcal{N}$, the expected payments are consistent. This definition of consistency is motivated by the following observations. The price distortion is factored in the expected payments. From \eqref{eq:expectedpayments} we see that price distortions (premia) can bias benefits toward a subset of players. In particular, if the premium at a given node is negative ($\mathcal{M}_n^{\pi}<0$), a supplier will not benefit from the day-ahead market but a consumer will. This situation can prevent suppliers from participating in day-ahead market. If $\mathcal{M}_n^{\pi}>0$, the oppostive holds true. This situation can prevent consumers from providing price-responsive demands. We can thus conclude that price consistency ensures payment consistency with respect to suppliers and consumers. In other words, $\mathcal{M}^{\pi}_n=0$ implies \eqref{eq:payconsistency}. 

\citet{kaye} argue that setting the day-ahead prices to the expected real-time prices (price consistency) is desirable because it effectively eliminates the day-ahead component of the market. Consequently, the market operates (in expectation) as a pure real-time market. This situation is desirable because it implies that the day-ahead market does not interfere with the incentives provided by real-time markets. This is particularly important for players that benefit from real-time market variability (such as peaking units and price-response demands). This also implies that the ISO does not give any preference to either risk-taking or risk-averse players. We also highlight that price consistency does not imply that premia do not exist; they can exist in each scenario but not in expectation. 

Deterministic formulations can yield persistent price premia that benefit a subset of players or that can be used for market manipulation. For instance, consider the case in which a wind farm forecast has the same mean but very different variance (uncertainty) for several consecutive days. If the expected forecast is used, the day-ahead prices will be consistently the same for all days, thus making them more predictable and biased toward a subset of players. While the use of risk-adaptive reserves can help ameliorate this effect, this approach is not guaranteed to achieve price consistency. 

\subsection{Uplift Payments}

From \eqref{eq:expectedpayments} we see that if the premium at a given node is negative ($\mathcal{M}_n^{\pi}<0$), negative payments (losses) can be incurred by the suppliers. This issue is analyzed by \citet{fullerpricing} and \citet{morales2012pricing}. For instance, a wind supplier might be cleared at a given forecast capacity and at a low price but in real-time it might need to buy back power at a larger price if the realized capacity is lower than forecasted (this is illustrated in Section \ref{sec:computation}).  It is thus desired that suppliers be paid at least as much as what they asked for and it is desired that consumers do not pay more than what they are willing to pay for. This is formally stated in the following definition.

\begin{definition}[Wholeness]\label{def:wholeness}
We say that suppliers and consumers are {\em whole in expectation} if
\begin{subequations}
\begin{align}
\mathbb{E}\left[P_{i}^g(\omega)\right]&\geq \;\;\;\;\mathbb{E}\left[C_{i}^g(\omega)\right], \quad i\in\mathcal{G}\\
-\mathbb{E}\left[P_{j}^d(\omega)\right]&\leq -\mathbb{E}\left[C_{j}^d(\omega)\right], \quad j\in\mathcal{D}.
\end{align}
\end{subequations}
\end{definition}
If the players are not made whole, they can leave the market and this can hinder diversification. Uplift payments are routinely used by the ISOs to avoid this situation \citep{galianasocial,baldickuplift}. Uplifts can result from inadequate representations of system behavior such as nonconvexities \citep{oneill} or, as we will see in Section \ref{sec:computation}, can result from using inadequate statistical representations of real-time market performance in deterministic settings. Consequently, uplift payments are a useful metric to determine the effectiveness of a given clearing formulation. 
\begin{definition}[Uplift Payments]\label{def:uplift}
We define the {\em expected uplift payments} to suppliers and consumers as
\begin{subequations}\label{eq:uplift}
\begin{align}
\mathcal{M}^U_{i} &:= -\min\left\{\mathbb{E}[P^g_i(\omega)]-\mathbb{E}[C^g_i(\omega)],0\right\}, \quad i\in\mathcal{G}\\
\mathcal{M}^U_{j} &:= -\min\left\{\mathbb{E}[P^d_j(\omega)]-\mathbb{E}[C^d_j(\omega)],0\right\}, \quad j\in\mathcal{D}.
\end{align}
\end{subequations}
We also define the total uplift as $\displaystyle \mathcal{M}^U := \sum_{i\in\mathcal{G}} \mathcal{M}^U_{i}+\sum_{j\in\mathcal{D}} \mathcal{M}^U_{j}$. 
\end{definition}

We highlight that our setting is convex and we thus only consider uplifts arising from inadequate statistical representations.

\subsection{Revenue Adequacy}

An efficient clearing procedure must ensure that the ISO does not run into financial deficit. In other words, the ISO must have a positive cash flow (payments collected from consumers are greater than the payments given to suppliers). We consider the following expected revenue definition, used by \citet{philpott}, to assess performance with respect to this case.

\begin{definition}[Revenue Adequacy] The {\em expected net payment to the ISO} is defined as
\begin{align}\label{eq:revaqmetric}
\mathcal{M}^{ISO}&:=\mathbb{E}\left[\sum_{i \in\mathcal{G}}P_{i}^g(\omega)+\sum_{j \in\mathcal{D}}P_{j}^d(\omega)\right]\nonumber\\
&\;=\sum_{i\in\mathcal{G}}\mathbb{E}[P^g_i(\omega)]+\sum_{j\in\mathcal{D}}\mathbb{E}[P^d_j(\omega)].
\end{align}
We say that the ISO is {\em revenue adequate in expectation} if $\mathcal{M}^{ISO} \leq 0$.
\end{definition}

Revenue adequacy guarantees that, in expectation, the ISO will not run into financial deficit.  
\section{Properties of Stochastic Clearing}\label{sec:properties}

In this section, we prove that the stochastic clearing formulation yields bounded price distortions and that these distortions can be made arbitrarily small. In addition, we prove that day-ahead quantities are bounded by real-time quantities and that they converge to a quantile of the real-time quantities when the distortions are zero. Further, we prove that the formulation yields revenue adequacy and zero uplifts in expectation.

\subsection{No Network Constraints}\label{sec:prop:withoutnetwork}

We begin our discussion with a single-node formulation (no network constraints) and then generalize the results to the case of network constraints. The single-node formulation has the form:
\begin{subequations}\label{eq:stochnonet}
\begin{align}
  \min_{d_j,g_i,G_i(\cdot),D(\cdot)}\; 
  &\; \sum_{i\in\mathcal{G}} \mathbb{E}\left[ \alpha_i^g G_i(\omega) + \Delta\alpha_i^{g,+} (G_i(\omega) - g_i)_+ + \Delta\alpha_i^{g,-} (G_i(\omega) - g_i)_- \right] \notag \\
  & + \sum_{j\in\mathcal{D}} \mathbb{E}\left[ -\alpha_j^d D_j(\omega) + \Delta\alpha_j^{d,+} (D_j(\omega) - d_i)_- + \Delta\alpha_j^{d,-} (D_j(\omega) - d_j)_+ \right] \\
\textrm{s.t.} & \sum_{i\in\mathcal{G}} g_i = \sum_{j\in\mathcal{D}}d_j \quad (\pi)\\
              & \sum_{i\in\mathcal{G}} (G_i(\omega)-g_i) = \sum_{j\in\mathcal{D}}(D_j(\omega)-d_j) \quad \omega \in \Omega \quad (p(\omega)\Pi(\omega))\\
& 0\leq G_i(\omega)\leq \bar{G}_i(\omega),\quad i\in\mathcal{G},\omega \in \Omega\label{eq:bound1}\\
& 0\leq D_j(\omega)\leq \bar{D}(\omega),\quad j\in\mathcal{D},\omega \in \Omega\label{eq:bound2}.
\end{align}
\end{subequations}
This formulation assumes infinite transmission capacity. In this case, the entire network collapses into a single node; consequently, a single day-ahead price $\pi$ and real-time price $\Pi(\omega)$ are used.  

We state that the partial Lagrange function of \eqref{eq:stochnonet} is given by
\begin{align*}
  &\mathcal{L}(d_j,D_j(\cdot),g_i,G_i(\cdot),\pi,\Pi(\cdot)) \notag \\
  & = \sum_{i\in\mathcal{G}} \mathbb{E}\left[ \alpha_i^g G_i(\omega) + \Delta\alpha_i^{g,+} (G_i(\omega) - g_i)_+ + \Delta\alpha_i^{g,-} (G_i(\omega) - g_i)_- \right] \notag\\ 
  & \quad - \sum_{j\in\mathcal{D}} \mathbb{E}\left[ \alpha_j^d D_j(\omega) - \Delta\alpha_j^{d,+} (D_j(\omega) - d_i)_- - \Delta\alpha_j^{d,-} (D_j(\omega) - d_j)_+ \right] \notag \\
  & \quad - \pi\left(\sum_{i\in\mathcal{G}}g_i-\sum_{j\in\mathcal{D}}d_j\right) - \mathbb{E}\left[\Pi(\omega)\left(\sum_{i\in\mathcal{G}}(G_i(\omega)-g_i)-\sum_{j\in\mathcal{D}}(D_j(\omega)-d_j)\right)\right] .
\end{align*}
The contribution of the balance constraints can be written in expected value form if we weight the Lagrange multipliers of the balance equations (prices) by the probabilities $p(\omega)$. 

\begin{theorem}\label{th:singledistortion}
Consider the single-node stochastic clearing problem \eqref{eq:stochnonet}, and assume that the incremental bid prices are positive. The price distortion $\mathcal{M}^{\pi}=\pi-\mathbb{E}[\Pi(\omega)]$ is bounded as
\begin{align}
  -\Delta\alpha^+ \leq \mathcal{M}^{\pi}\leq \Delta\alpha^-,
\end{align}
where
\begin{subequations}\label{eq:delta}
\begin{align}\label{eq:deltaplus}
  \Delta \alpha^+ = \min \left\{\min_{i\in\mathcal{G}} \Delta\alpha_i^{g,+}, \min_{j\in\mathcal{D}} \Delta\alpha_j^{d,+} \right\}
\end{align}
and
\begin{align}\label{eq:deltaminus}
  \Delta \alpha^- = \min \left\{\min_{i\in\mathcal{G}} \Delta\alpha_i^{g,-}, \min_{j\in\mathcal{D}} \Delta\alpha_j^{d,-} \right\}.
\end{align}
\end{subequations}
\end{theorem}
\proof{Proof}
Since $(X-x)_- = (X-x)_+ - (X-x)$, we have the partial Lagrange function
\begin{align*}
  &\mathcal{L}(d_j,D_j(\cdot),g_i,G_i(\cdot),\pi,\Pi(\cdot)) \notag \\
  & = \sum_{i\in\mathcal{G}} \mathbb{E}\left[ \alpha_i^g G_i(\omega) + \Delta\alpha_i^{g,+} (G_i(\omega) - g_i)_+ + \Delta\alpha_i^{g,-} (G_i(\omega) - g_i)_+ - \Delta\alpha_i^{g,-} (G_i(\omega) - g_i) \right] \notag\\ 
  & \quad - \sum_{j\in\mathcal{D}} \mathbb{E}\left[ \alpha_j^d D_j(\omega) - \Delta\alpha_j^{d,+} (D_j(\omega) - d_i)_+ + \Delta\alpha_j^{d,+} (D_j(\omega) - d_i) - \Delta\alpha_j^{d,-} (D_j(\omega) - d_j)_+ \right] \notag \\
  & \quad - \pi\left(\sum_{i\in\mathcal{G}}g_i-\sum_{j\in\mathcal{D}}d_j\right) - \mathbb{E}\left[\Pi(\omega)\left(\sum_{i\in\mathcal{G}}(G_i(\omega)-g_i)-\sum_{j\in\mathcal{D}}(D_j(\omega)-d_j)\right)\right].
\end{align*}
The stationarity conditions of the partial Lagrange function with respect to the day-ahead quantities $d_j,g_i$ are given by
\begin{subequations}
\begin{align}
  0 &\in \partial_{d_j}\mathcal{L} 
  = (\Delta\alpha_j^{d,+} + \Delta\alpha_j^{d,-}) \partial_{d_j} \mathbb{E}[(D_j(\omega) - d_j)_+] + \Delta\alpha_j^{d,+} + \pi - \mathbb{E}[\Pi(\omega)] \quad j\in\mathcal{D} \label{eq:stationd}\\
  0 &\in \partial_{g_i}\mathcal{L}
  = (\Delta\alpha_i^{g,+} + \Delta\alpha_i^{g,-}) \partial_{g_i} \mathbb{E}[(G_i(\omega) - g_i)_+] + \Delta\alpha_i^{g,-} - \pi + \mathbb{E}[\Pi(\omega)] \quad i\in\mathcal{G}. \label{eq:stationg}
\end{align}
\end{subequations}
Rearranging \eqref{eq:stationd}, we obtain
\begin{subequations}\label{eq:stations}
\begin{align}
  \frac{- \Delta\alpha_j^{d,+} - \pi + \mathbb{E}[\Pi(\omega)]}{\Delta\alpha_j^{d,+} + \Delta\alpha_j^{d,-}} 
  & \in \partial_{d_j} \mathbb{E}[(D_j(\omega) - d_j)_+] \quad j\in\mathcal{D} \label{eq:stations:d}\\
  \frac{- \Delta\alpha_i^{g,-} + \pi - \mathbb{E}[\Pi(\omega)]}{\Delta\alpha_i^{g,+} + \Delta\alpha_i^{g,-}} 
  & \in \partial_{g_i} \mathbb{E}[(G_i(\omega) - g_i)_+] \quad i\in\mathcal{G}.
\end{align}
\end{subequations}

From the property
\begin{align}
\label{eq:subdiff}
  \partial_x (X-x)_+ = \begin{cases}
    -1 & \text{if } X > x \\
     0 & \text{if } X < x \\
    [-1, 0] & \text{if } X = x
  \end{cases},
\end{align} 
we have
\begin{align}
  \partial_{x}\mathbb{E}[(X(\omega)-x)_+] \notag
  &= \left\{\mathbb{E}\left[-\mathbf{1}_{X(\omega) > x} + a\mathbf{1}_{X(\omega) = x}\right] \;:\; a \in [-1,0] \right\} \notag \\
  &= \left\{-\mathbb{P}\left(X(\omega) > x\right) + a\mathbb{P}\left(X(\omega) = x\right) \;:\; a \in [-1,0] \right\}, \notag\\
  &= \left\{\eta \;:\; -\mathbb{P}(X(\omega) \geq x) \leq \eta \leq -\mathbb{P}(X(\omega) > x) \right\}. \label{eq:probabilityrange}
\end{align}
Since $-1\leq -\mathbb{P}(X(\omega) \geq x) \leq -\mathbb{P}(X(\omega) > x) \leq 0$, we have
\begin{align}\label{eq:subdifferential}
  \partial_{x}\mathbb{E}[(X(\omega)-x)_+] \subseteq [-1,0].
\end{align}

From \eqref{eq:stations} and \eqref{eq:subdifferential}, we have
\begin{subequations}\label{eq:subgradbound}
\begin{align}
  -1 \leq \frac{-\Delta\alpha_j^{d,+} - \pi + \mathbb{E}\left[\Pi(\omega)\right]}{\Delta\alpha_j^{d,+} + \Delta\alpha_j^{d,-}} \leq 0 \label{eq:subgradbounddj}\\
  -1 \leq \frac{-\Delta\alpha_i^{g,-} + \pi - \mathbb{E}\left[\Pi(\omega)\right]}{\Delta\alpha_i^{g,+} + \Delta\alpha_i^{g,-}} \leq 0.\label{eq:subgradboundgi}
\end{align}
\end{subequations}
The above relationships are equivalent to
\begin{subequations}\label{eq:distbound}
\begin{align}
  -\Delta\alpha_j^{d,+} &\leq \pi-\mathbb{E}\left[\Pi(\omega)\right] \leq \Delta\alpha_j^{d,-} \\
  -\Delta\alpha_i^{g,+} & \leq \pi-\mathbb{E}\left[\Pi(\omega)\right] \leq \Delta\alpha_i^{g,-}.
\end{align}
\end{subequations}
or, equivalently, $-\Delta\alpha^+ \leq \mathcal{M}^{\pi}\leq \Delta\alpha^-$.\Halmos
\endproof

The price distortion is bounded above by the smallest of all $\Delta\alpha^{g,-}_i$ and $\Delta\alpha^{d,-}_j$ and bounded below by the largest of all $-\Delta\alpha^{g,+}_i$ and $-\Delta\alpha^{d,+}_j$. The bounds are denoted by $\Delta\alpha^-$ and $-\Delta\alpha^+$, respectively. This implies that if we let $\Delta\alpha^+$ and $\Delta\alpha^-$ be sufficiently small, then {\em we can make the price distortion $\mathcal{M}^{\pi}$ arbitrarily small}. Note that the bound is independent of the cleared quantities, which reflects robust behavior. Moreover, the upper bound depends on the incremental bid prices $\Delta\alpha_i^{g,-}$ and $\Delta\alpha_j^{d,-}$ only, while the lower bound depends on $\Delta\alpha_i^{g,+}$ and $\Delta\alpha_j^{d,+}$ only.  Boundedness of the price distortion also eliminates the day-ahead component of the suppliers and consumer payments and thus achieves payment consistency. We highlight that Theorem~\ref{th:singledistortion} assumes the positive incremental bid prices. Otherwise, the solution can be degenerate.

We now prove that the day-ahead quantities $d_j,g_i$ obtained from the stochastic clearing model are implicitly bounded by the minimum and maximum real-time quantities.

\begin{theorem}\label{th:singlebound}
Consider the single-node stochastic clearing problem \eqref{eq:stochnonet}, and assume that the incremental bid prices are positive. The day-ahead quantities are bounded by the real-time quantities as
\begin{align*}
  \min_{\omega \in \Omega}D_j(\omega)\leq d_j\leq \max_{\omega \in \Omega}\,D_j(\omega), \; j\in\mathcal{D}\\
  \min_{\omega \in \Omega}G_i(\omega)\leq g_i\leq \max_{\omega \in \Omega}\,G_i(\omega), \; i\in\mathcal{G}.
\end{align*}
\end{theorem}
\proof{Proof}
Consider the following two cases:
\begin{itemize}
  \item Case 1: The price distortion hits the lower bound for demand $j$; we thus have  $\pi-\mathbb{E}\left[\Pi(\omega)\right]=-\Delta\alpha^{d,+}_j$. This implies $0 \in \partial_{d_j} \mathbb{E}[(D_j(\omega) - d_j)_+]$ from \eqref{eq:stations:d}, and hence $\mathbb{P}(D_j(\omega) > d_j) = 0$ and $\mathbb{P}(D_j(\omega) \leq d_j) = 1$ from \eqref{eq:subdiff}. This implies that $d_j\geq D_j(\omega),\; \forall \omega \in \Omega$ and $d_j\geq \min_{\omega \in \Omega} {D}_j(\omega)$.
  \item Case 2: The price distortion hits the upper bound for demand $j$; we thus have $\pi-\mathbb{E}\left[\Pi(\omega)\right]=\Delta\alpha^{d,-}_j$. This implies $-1 \in \partial_{d_j} \mathbb{E}[(D_j(\omega) - d_j)_+]$ from \eqref{eq:stations:d}, and hence $\mathbb{P}(D_j(\omega) \geq d_j) = 1$ from \eqref{eq:subdiff}. This implies that $d_j\leq D_j(\omega),\; \forall \omega \in \Omega$ and $d_j \leq \max_{\omega \in \Omega}D_j(\omega)$.
\end{itemize}

We thus conclude that $d_j$ is bounded from below by $\min_{\omega \in \Omega}D_j(\omega)$ and from above by $\max_{\omega \in \Omega}D_j(\omega)$. The same procedure can be followed to prove that $g_i$ is bounded from below by $\min_{\omega \in \Omega}G_i(\omega)$ and from above by $\max_{\omega \in \Omega}G_i(\omega)$.\Halmos
\endproof

The implicit bound on the day-ahead quantities $d_j,g_i$ is a key property of the stochastic model proposed because it implies that we do not have to choose day-ahead capacities $\bar{g}_i,\bar{d}_j$ (e.g., summarization statistics). These are automatically set by the model through the scenario information. This is important because, as we have mentioned, obtaining proper summarizing statistics for complex probability distributions might not be trivial.

We now prove that if the price distortion is zero, the day-ahead quantities converge to {\em quantiles} of the real-time quantities.

\begin{theorem}\label{th:singlemedian}
Consider the stochastic clearing problem \eqref{eq:stochnonet}, and assume that the incremental bid prices are positive. If the price distortion is zero at the solution, then
\begin{subequations}
\begin{align}
  d_j &= \mathbb{Q}_{D_j(\omega)}\left( \frac{\Delta\alpha_j^{d,-}}{\Delta\alpha_j^{d,+} + \Delta\alpha_j^{d,-}} \right), \; j\in\mathcal{D} \label{eq:quantiledj}\\ 
  g_i &= \mathbb{Q}_{G_i(\omega)}\left( \frac{\Delta\alpha_i^{g,+}}{\Delta\alpha_i^{g,+} + \Delta\alpha_i^{g,-}} \right), \; i \in\mathcal{G}.\label{eq:quantilegi}
\end{align}
\end{subequations}
\end{theorem}

\proof{Proof}
From \eqref{eq:probabilityrange} and \eqref{eq:subgradbounddj} we have that if $\pi - \mathbb{E}\left[\Pi(\omega)\right]=0$, then $-\mathbb{P}(D_j(\omega) \geq d_j) \leq \frac{-\Delta\alpha_j^{d,+}}{\Delta\alpha_j^{d,+} + \Delta\alpha_j^{d,-}}\leq -\mathbb{P}(D_j(\omega) > d_j)$, and thus $\mathbb{P}(D_j(\omega) < d_j) \leq \frac{\Delta\alpha_j^{d,-}}{\Delta\alpha_j^{d,+} + \Delta\alpha_j^{d,-}}\leq \mathbb{P}(D_j(\omega) \leq d_j)$. This implies \eqref{eq:quantiledj} from \eqref{eq:defQuantile}. The same argument holds for \eqref{eq:quantilegi}.\Halmos
\endproof

\begin{corollary}
If the incremental bid prices are symmetric then $d_j = \mathbb{M}\left(D_j(\omega)\right),\; j\in\mathcal{D}$ and $g_i = \mathbb{M}\left(D_j(\omega)\right),\; i\in\mathcal{G}$.
\end{corollary}
\proof{Proof}
The proof follows from Corollary~\ref{thm:abscost} and Theorem \ref{th:singlemedian}. \Halmos
\endproof

This result implies that the day-ahead quantities $d_j,g_i$ cannot in general be guaranteed to converge to the expected values of the real-time quantities $\mathbb{E}[D_j(\omega)],\mathbb{E}[G_i(\omega)]$. Such convergence can only be guaranteed when the quantile and mean coincide. These observations thus imply that {\em the expected value is not necessarily the only statistic that can be used for the capacities in the day-ahead market.}

We now prove that the stochastic formulation yields zero uplifts in expectation. Revenue adequacy is not considered because this is a single-node problem. We use the strategy followed by \citet{morales2012pricing}. For this discussion, we denote a minimizer of the partial Lagrange function (subject to the constraints \eqref{eq:bound1} and \eqref{eq:bound2}) as $d_j^*,D_j(\cdot)^*,g_i^*,G_i(\cdot)^*,\pi^*,\Pi^*(\cdot)$. Because the problem is convex, we know that the optimal prices $\pi^*,\Pi^*(\cdot)$ satisfy
\begin{align}\label{eq:singlelag}
(d_j^*,D_j(\cdot)^*,g_i^*,G_i^*(\cdot)) = \mathop{\textrm{argmin}}_{d_j,D_j(\cdot),g_i,G_i(\cdot)}  & \quad \mathcal{L}(d_j,D_j(\cdot),g_i,G_i(\cdot),\pi^*,\Pi^*(\cdot))\quad \textrm{s.t.} \quad  \eqref{eq:bound1}-\eqref{eq:bound2}.
\end{align}
Moreover, at fixed $\pi^*,\Pi^*(\cdot)$, the partial Lagrange function can be 
separated as
\begin{align}
\mathcal{L}(d_j,D_j(\cdot),g_i,G_i(\cdot),\pi^*,\Pi^*(\cdot))=\sum_{i \in \mathcal{G}}\mathcal{L}_i^g(g_i,G_i(\cdot),\pi^*,\Pi^*(\cdot))+\sum_{j \in \mathcal{D}}\mathcal{L}_j^d(d_j,D_j(\cdot),\pi^*,\Pi^*(\cdot)),
\end{align}
where
\begin{subequations}\label{eq:contlag}
\begin{align}
\mathcal{L}_i^g(g_i,G_i(\cdot),\pi^*,\Pi^*(\cdot))& := \mathbb{E}[C_i^g(\omega)]-\mathbb{E}[P_i^g(\omega)],\; i\in \mathcal{G}\\
\mathcal{L}_j^d(d_j,D_j(\cdot),\pi^*,\Pi^*(\cdot))& := \mathbb{E}[C_j^d(\omega)]-\mathbb{E}[P_j^d(\omega)],\; j\in\mathcal{D}.
\end{align}
\end{subequations}
Consequently, one can minimize the partial Lagrange function by minimizing \eqref{eq:contlag} independently. 

\begin{theorem} \label{thm:zerouplift}
Consider the single-node clearing problem \eqref{eq:stochnonet}, and let the assumptions of Theorem \ref{th:singledistortion} hold. Any minimizer $d_j^*,D_j(\cdot)^*,g_i^*,G_i(\cdot)^*,\pi^*,\Pi^*(\cdot)$ of \eqref{eq:stochnonet} yields zero uplift payments in expectation:
\begin{subequations}
\begin{align}
\mathcal{M}_i^U&=0, \quad i\in \mathcal{G}\\ 
\mathcal{M}_j^U&=0, \quad j\in \mathcal{D}. 
\end{align}
\end{subequations}
\end{theorem}

\proof{Proof}
From Definition~\ref{def:uplift}, it suffices to show that $\mathbb{E}[P_i^g(\omega)] - \mathbb{E}[C_i^g(\omega)] \geq 0$ for all $i\in \mathcal{G}$ and $\mathbb{E}[P_j^d(\omega)] - \mathbb{E}[C_j^d(\omega)]\geq 0$ for all $j\in \mathcal{D}$. For fixed $\pi^*,\Pi^*(\omega)$, the candidate solution $d_j=D_j(\cdot)=g_i=G_i(\cdot)=0$ is feasible for \eqref{eq:singlelag} with values $\mathcal{L}^g_i(g_i,G_i(\cdot),\pi^*,\Pi^*(\cdot))=0,\; i\in\mathcal{G}$ and $\mathcal{L}^d(d_j,D_j(\cdot),\pi^*,\Pi^*(\cdot))=0,\; j\in\mathcal{D}$. Because the candidate is suboptimal we have $\mathcal{L}_i^g(g_i^*,G_i^*(\cdot),\pi^*,\Pi^*(\cdot)) \leq  \mathcal{L}_i^g(g_i,G_i(\cdot),\pi^*,\Pi^*(\cdot))=0$ and $\mathcal{L}_j^d(d_j^*,D_j^*(\cdot),\pi^*,\Pi^*(\cdot)) \leq 0$. The result follows from equations \eqref{eq:contlag} and the definition of $\mathcal{M}_i^U$ and $\mathcal{M}_j^U$ in \eqref{eq:uplift}.\Halmos
\endproof

\subsection{Network Constraints}\label{sec:prop:withnetwork}

Having established some insights into the properties of the stochastic model, we now turn our attention to the full stochastic problem with network constraints \eqref{eq:stoch} and generalize our results.  

It is well known that stochastic formulations yield a better expected social surplus. This follows from the well-known inequality (see \citet{birge}):
\begin{align}
\varphi^{sto}_{WS}\leq \varphi^{sto}\leq \varphi^{det}.
\end{align}
This follows from the fact that the stochastic formulation will lead to a lower recourse cost (real-time penalty costs) than will the deterministic solution because the deterministic day-ahead problem does not anticipate recourse actions. The wait-and-see setting can perfectly anticipate real-time market conditions and therefore its real-time penalties are zero. This makes it the optimal, but nonimplementable policy. 

We now establish boundedness of the price distortions throughout the network. To establish our result, we need the following definitions. We rewrite equations \eqref{eq:DAangle} and \eqref{eq:RTangle} as
\begin{subequations}
\label{eq:angle2}
\begin{align}
  & f_\ell = \sum_{n\in\mathcal{N}} B_{\ell n} \theta_n \quad \forall \ell\in\mathcal{L}, \label{eq:DAangle2}\\
  & F_\ell(\omega) = \sum_{n\in\mathcal{N}} B_{\ell n} \Theta_n(\omega) \quad \forall \omega\in\Omega, \ell\in\mathcal{L},\label{eq:RTangle2}
\end{align}
\end{subequations}
where
\begin{align*}
  B_{\ell n} = \begin{cases}
    B_\ell & \text{if } n = rec(\ell), \\
    -B_\ell & \text{if } n = snd(\ell), \\
    0 & \text{otherwise}.
  \end{cases}
\end{align*}
We note that the above definitions imply that $B_{\ell, rec(\ell)}=B_\ell$ and $B_{\ell, snd(\ell)}=-B_\ell$. Moreover we have that,
\begin{align*}
\sum_{\ell\in \mathcal{L}_n^{rec}} f_\ell&=\sum_{\ell \in \mathcal{L}_n^{rec}} \left(B_{\ell}\theta_{rec(\ell)}-B_\ell\theta_{snd(\ell)}\right)\\
&=\sum_{\ell \in \mathcal{L}_n^{rec}}\left(B_{\ell, rec(\ell)}\theta_{rec(\ell)}+B_{\ell, snd(\ell)}\theta_{snd(\ell)}\right)\\
&=\sum_{\ell \in \mathcal{L}_n^{rec}}\left(B_{\ell n}\theta_{n}+B_{\ell, snd(\ell)}\theta_{snd(\ell)}\right).
\end{align*}
Using similar observations we have that,
\begin{align}
\sum_{\ell\in \mathcal{L}_n^{snd}} f_\ell=\sum_{\ell \in \mathcal{L}_n^{snd}} \left(B_{\ell,rec(\ell)} \theta_{rec(\ell)} + B_{\ell n} \theta_n\right).
\end{align}
Substituting the flows $f_\ell, F_\ell(\omega)$ by their corresponding phase angle expressions and using the above properties we have that the stochastic clearing problem \eqref{eq:angle2} can be written as
\newpage
\begin{subequations}
\label{eq:stochangle}
\begin{align}
  \min_{d_j,D_j(\cdot),g_i,G_i(\cdot),\theta_n,\Theta_n(\cdot)} \;
  & \sum_{i \in \mathcal{G}} \mathbb{E}\left[\alpha_i^{g} G_{i}(\omega) + \Delta\alpha_{i}^{g,+} (G_{i}(\omega)-g_{i})_+  + \Delta\alpha_{i}^{g,-} (G_{i}(\omega)-g_{i})_-\right]\nonumber\\
  & + \sum_{j \in \mathcal{D}} \mathbb{E}\left[-\alpha_{j}^{d} D_{j}(\omega) + \Delta\alpha_{j}^{d,+} (D_{j}(\omega) - d_{j})_- + \Delta\alpha_{j}^{d,-} (D_{j}(\omega) - d_{j})_+\right]\nonumber\\
  & + \sum_{\ell\in\mathcal{L}} \mathbb{E}\left[\Delta\alpha^{f,+}_{\ell} \left(\sum_{n\in\mathcal{N}} B_{\ell n} (\Theta_n(\omega) - \theta_n) \right)_+ + \Delta\alpha^{f,-}_{\ell} \left( \sum_{n\in\mathcal{N}} B_{\ell n} (\Theta_n(\omega) - \theta_n) \right)_-\right]\notag\\
  & + \sum_{n\in\mathcal{N}} \mathbb{E}\left[\Delta\alpha^{\theta,+}_n (\Theta_n(\omega) - \theta_n)_+ + \Delta\alpha^{\theta,-}_n (\Theta_n(\omega) - \theta_n)_-\right]\\
  \text{s.t.} 
  & \sum_{\ell \in \mathcal{L}_n^{rec}} \left(B_{\ell n} \theta_n + B_{\ell,snd(\ell)} \theta_{snd(\ell)}\right) -\sum_{\ell \in \mathcal{L}_n^{snd}} \left(B_{\ell,rec(\ell)} \theta_{rec(\ell)} + B_{\ell n} \theta_n\right) \notag \\
  & \quad + \sum_{i \in \mathcal{G}_n} g_{i} - \sum_{i \in \mathcal{D}_n} d_{i}=0,\quad (\pi_n) \quad \forall n\in\mathcal{N} \label{eq:ec:stochconst1} \\
  & \sum_{\ell \in \mathcal{L}_n^{rec}} \left[ B_{\ell n} \left( \Theta_n(\omega) - \theta_n \right) + B_{\ell,snd(\ell)} \left( \Theta_{snd(\ell)}(\omega) - \theta_{snd(\ell)} \right) \right] \notag \\
  & \quad -\sum_{\ell \in \mathcal{L}_n^{snd}} \left[ B_{\ell,rec(\ell)} \left( \Theta_{rec(\ell)}(\omega) - \theta_{rec(\ell)} \right) + B_{\ell n} \left( \Theta_n(\omega) - \theta_n \right) \right] \notag \\
  & \quad + \sum_{i \in \mathcal{G}_n} \left(G_{i}(\omega)-g_{i}\right) - \sum_{j \in \mathcal{D}_n} \left(D_{j}(\omega)-d_{j}\right)=0, \quad (p(\omega)\Pi_n(\omega)),\quad \forall\omega \in \Omega,n\in\mathcal{N} \label{eq:ec:stochconst2}\\
  & -\bar{F}_{\ell}(\omega)\leq \sum_{n\in\mathcal{N}} B_{\ell n} \Theta_n(\omega) \leq \bar{F}_{\ell}(\omega), \quad \forall\omega\in\Omega,\ell \in \mathcal{L} \label{eq:ec:stochconst3}\\
  & 0\leq G_{i}(\omega)\leq \bar{G}_{i}(\omega),\quad \forall\omega\in\Omega,i\in\mathcal{G} \\
  & 0\leq D_{j}(\omega)\leq \bar{D}_{j}(\omega),\quad \forall\omega\in\Omega,j\in\mathcal{D} \label{eq:ec:stochconst6}\\
  &\underline\theta_n \leq \Theta_n(\omega) \leq \overline\theta_n, \quad \forall\omega\in\Omega,n\in\mathcal{N}.
\end{align}
\end{subequations}
We consider the partial Lagrange function of \eqref{eq:stochangle} as
\begin{align}\label{eq:lagstoch}
  &\mathcal{L}(d_j,D_j(\cdot),g_i,G_i(\cdot),\theta_n,\Theta_n(\cdot),\pi_n,\Pi_n(\cdot)) \notag\\
  &= \sum_{i\in\mathcal{G}} \mathbb{E}\left[\alpha_i^g G_i(\omega) + \left(\Delta\alpha_i^{g,+} + \Delta\alpha_i^{g,-}\right)(G_{i}(\omega)-g_i)_+ - \Delta\alpha_i^{g,-}(G_{i}(\omega)-g_i)\right] \notag\\
  &\quad -\sum_{j\in\mathcal{D}} \mathbb{E}\left[\alpha^d_jD_j(\omega) - \left(\Delta\alpha^{d,+}_j+\Delta\alpha^{d,-}_j\right)(D_j(\omega)-d_j)_+ + \Delta\alpha^{d,+}_j(D_j(\omega)-d_j)\right] \notag\\
  &\quad + \sum_{\ell\in\mathcal{L}} \mathbb{E}\left[\left(\Delta\alpha^{f,+}_{\ell}+\Delta\alpha^{f,-}_{\ell}\right) \left( \sum_{n\in\mathcal{N}} B_{\ell n} (\Theta_n(\omega) - \theta_n) \right)_+ - \Delta\alpha^{f,-}_{\ell} \sum_{n\in\mathcal{N}} B_{\ell n} (\Theta_n(\omega) - \theta_n) \right] \notag\\
  &\quad + \sum_{n\in\mathcal{N}} \mathbb{E}\left[ \left(\Delta\alpha^{\theta,+}_n + \Delta\alpha^{\theta,-}_n \right) \left(\Theta_n(\omega) - \theta_n \right)_+ - \Delta\alpha^{\theta,-}_n \left(\Theta_n(\omega) - \theta_n\right)\right]\notag\\
  &\quad -\sum_{n\in\mathcal{N}}\pi_n \left[ \sum_{\ell \in \mathcal{L}_n^{rec}} \left(B_{\ell n} \theta_n + B_{\ell,snd(\ell)} \theta_{snd(\ell)}\right) -\sum_{\ell \in \mathcal{L}_n^{snd}} \left(B_{\ell,rec(\ell)} \theta_{rec(\ell)} + B_{\ell n} \theta_n\right) + \sum_{i \in \mathcal{G}_n} g_{i} - \sum_{j \in \mathcal{D}_n} d_j \right] \notag\\
  &\quad -\mathbb{E} \left[ \sum_{n\in\mathcal{N}} \Pi_n(\omega) \left(\sum_{\ell \in \mathcal{L}_n^{rec}} \left[ B_{\ell n} \left( \Theta_n(\omega) - \theta_n \right) + B_{\ell,snd(\ell)} \left( \Theta_{snd(\ell)}(\omega) - \theta_{snd(\ell)} \right) \right] \right.\right. \notag \\
  & \qquad \qquad \qquad \qquad -\sum_{\ell \in \mathcal{L}_n^{snd}} \left[ B_{\ell,rec(\ell)} \left( \Theta_{rec(\ell)}(\omega) - \theta_{rec(\ell)} \right) + B_{\ell n} \left( \Theta_n(\omega) - \theta_n \right) \right] \notag \\
  & \qquad \qquad \qquad \qquad \left.\left. + \sum_{i \in \mathcal{G}_n} \left(G_{i}(\omega)-g_{i}\right) - \sum_{j \in \mathcal{D}_n} \left(D_{j}(\omega)-d_{j}\right) \right)\right].
\end{align}
We define the subset $\bar{\mathcal{N}}\subseteq \mathcal{N}$ containing all nodes at which at least one supplier or consumer is connected. We also define the subset $\mathcal{L}_n := \mathcal{L}_n^{rec} \cup \mathcal{L}_n^{snd}$.

\begin{theorem}\label{th:networkdistortion}
Consider the stochastic clearing model \eqref{eq:stochangle} and assume that the incremental bid prices are positive and that $\Delta\alpha^{f,+}_{\ell}, \Delta\alpha^{f,-}_{\ell},\Delta\alpha^{\theta,+}_n,\Delta\alpha^{\theta,-}_n>0,\; \ell\in \mathcal{L}, \; n\in\mathcal{N}$. The price distortions $\mathcal{M}_n^{\pi},\; n\in\mathcal{N}$ are bounded as
\begin{align*}
  -\Delta\bar\alpha_n^+ & \leq \mathcal{M}_n^{\pi} \leq \Delta\bar\alpha_n^-,\quad n\in\bar{\mathcal{N}}, \\
  -\Delta\alpha_n^+ & \leq \mathcal{M}_n^\pi \leq \Delta\alpha_n^-,\quad n\in\mathcal{N}\backslash\bar{\mathcal{N}},
\end{align*}
where
\begin{align*}
  \Delta\bar\alpha_n^+ &= \min\left\{\min_{i\in\mathcal{G}_n}\Delta\alpha_i^{g,+}, \min_{j\in\mathcal{D}_n}\Delta\alpha_j^{d,+}, \Delta\alpha_n^{+} \right\},\; n\in\bar{\mathcal{N}}, \\
  \Delta\bar\alpha_n^- &= \min\left\{\min_{i\in\mathcal{G}_n}\Delta\alpha_i^{g,-}, \min_{j\in\mathcal{D}_n} \Delta\alpha_j^{d,-}, \Delta\alpha_n^{-}\right\},\; n\in\bar{\mathcal{N}}
\end{align*}
and
\begin{align*}
  \Delta\alpha_n^{+} &= \displaystyle\frac{\sum_{\ell\in\mathcal{L}_n} \left( \Delta\alpha^{f,+}_{\ell} + (1 - B_{\ell n}) \Delta\alpha^{f,-}_\ell \right) + \Delta\alpha_n^{\theta,+}}{\sum_{\ell\in\mathcal{L}_n} B_{\ell}}, \; n\in\mathcal{N}, \\
  \Delta\alpha_n^{-} &=\displaystyle \frac{\sum_{\ell\in\mathcal{L}_n} B_{\ell n} \Delta\alpha^{f,-}_{\ell} + \Delta\alpha_n^{\theta,-}}{\sum_{\ell\in\mathcal{L}_n} B_{\ell}}, \; n\in\mathcal{N}.
\end{align*}
\end{theorem}

\proof{Proof}
The stationarity conditions of the partial Lagrange function with respect to the day-ahead quantities $g_i,d_j$ and phase angles $\theta_n$ are given by
\begin{subequations}\label{eq:stationd2}
\begin{align}
  0 &\in \partial_{d_j}\mathcal{L} 
  = (\Delta\alpha^{d,+}_j + \Delta\alpha^{d,-}_j) \partial_{d_j} \mathbb{E}[(D_j(\omega) - d_j)_+] + \Delta\alpha^{d,+}_j + \pi_{n(j)} - \mathbb{E}\left[\Pi_{n(j)}(\omega)\right] \quad j\in\mathcal{D}\\
  0 &\in \partial_{g_i}\mathcal{L} 
  = (\Delta\alpha^{g,+}_i + \Delta\alpha^{g,-}_i) \partial_{g_i} \mathbb{E}[(G_i(\omega) - g_i)_+] + \Delta\alpha^{g,-}_i - \pi_{n(i)} + \mathbb{E}\left[\Pi_{n(i)}(\omega)\right] \quad i\in\mathcal{G}\\
  0 &\in \partial_{\theta_n}\mathcal{L} 
  = \sum_{\ell\in\mathcal{L}_n} \left(\Delta\alpha^{f,+}_\ell + \Delta\alpha^{f,-}_\ell\right) \partial_{\theta_n} \mathbb{E}\left[\left(\sum_{m\in\mathcal{N}} B_{\ell m}(\Theta_m(\omega) - \theta_m) \right)_+\right] + \sum_{\ell\in\mathcal{L}_n} B_{\ell n} \Delta\alpha^{f,-}_\ell \notag \\
  & \qquad \qquad + \left( \Delta\alpha_n^{\theta,+} + \Delta\alpha_n^{\theta,-} \right) \partial_{\theta_n} \mathbb{E} \left[ \left( \Theta_n(\omega) - \theta_n \right)_+ \right] + \Delta\alpha_n^{\theta,-} \notag \\
  & \qquad \qquad - \left( \sum_{\ell\in\mathcal{L}_n^{rec}} B_{\ell n} - \sum_{\ell\in\mathcal{L}_n^{snd}} B_{\ell n} \right) \left( \pi_n - \mathbb{E}\left[\Pi_n(\omega)\right] \right) \quad n\in\mathcal{N}, \label{eq:stationd:angle}
\end{align}
\end{subequations}
where we recall that $n(i)$ is the node at which supplier $i$ is connected and $n(j)$ is the node at which demand $j$ is connected. Following the same bounding procedure used in the proof of Theorem \ref{th:singledistortion}, we obtain
\begin{align*}
  -\Delta\alpha_j^{d,+}\leq \mathcal{M}^{\pi}_{n(j)}\leq \Delta \alpha_j^{d,-},\;j\in\mathcal{D}\\
  -\Delta\alpha_i^{g,+}\leq \mathcal{M}^{\pi}_{n(i)}\leq \Delta \alpha_i^{g,-},\;i\in\mathcal{G}.
\end{align*}
Rearranging \eqref{eq:stationd:angle}, we obtain
\begin{align}\label{eq:subdiff:angle}
  -\sum_{\ell\in\mathcal{L}_n} B_{\ell n} \Delta\alpha^{f,-}_{\ell} - \Delta\alpha_n^{\theta,-} + \left( \sum_{\ell\in\mathcal{L}_n^{rec}} B_{\ell n} - \sum_{\ell\in\mathcal{L}_n^{snd}} B_{\ell n} \right) \mathcal{M}^{\pi}_n \in \mathcal{S}_n,
\end{align}
where
\begin{align}
  \mathcal{S}_n := 
  & \sum_{\ell\in\mathcal{L}_n} \left( \Delta\alpha^{f,+}_{\ell} + \Delta\alpha^{f,-}_{\ell} \right)\partial_{\theta_n} \mathbb{E}\left[\left(\sum_{m\in\mathcal{N}} B_{\ell m}(\Theta_m(\omega) - \theta_m) \right)_+\right] \notag\\
  & +  \left( \Delta\alpha_n^{\theta,+} + \Delta\alpha_n^{\theta,-} \right) \partial_{\theta_n} \mathbb{E} \left[ \left( \Theta_n(\omega) - \theta_n \right)_+ \right] \label{eq:sumofsubdiffs}
\end{align}
Because $\partial_{\theta_n} \mathbb{E}\left[\left(\sum_{m\in\mathcal{N}} B_{\ell m}(\Theta_m(\omega) - \theta_m) \right)_+\right] \subseteq [-1,0]$ and $\partial_{\theta_n} \mathbb{E} \left[ \left( \Theta_n(\omega) - \theta_n \right)_+ \right] \subseteq \left[ -1, 0 \right]$, we have
\begin{align}
  \mathcal{S}_n 
  \subseteq \left[ -\sum_{\ell\in\mathcal{L}_n} \left( \Delta\alpha^{f,+}_{\ell} + \Delta\alpha^{f,-}_{\ell} \right) - \Delta\alpha_n^{\theta,+} - \Delta\alpha_n^{\theta,-}, 0 \right], \label{eq:anglesubdiffrange}
\end{align}
and therefore, from \eqref{eq:subdiff:angle} and \eqref{eq:anglesubdiffrange},
\begin{align}
  & -\sum_{\ell\in\mathcal{L}_n} \left( \Delta\alpha^{f,+}_{\ell} + \Delta\alpha^{f,-}_{\ell} \right) - \Delta\alpha_n^{\theta,+} - \Delta\alpha_n^{\theta,-} \notag\\
  & \qquad \leq -\sum_{\ell\in\mathcal{L}_n} B_{\ell n} \Delta\alpha^{f,-}_{\ell} - \Delta\alpha_n^{\theta,-} + \left( \sum_{\ell\in\mathcal{L}_n^{rec}} B_{\ell n} - \sum_{\ell\in\mathcal{L}_n^{snd}} B_{\ell n} \right) \mathcal{M}^{\pi}_n \notag \\
  & \qquad = -\sum_{\ell\in\mathcal{L}_n} B_{\ell n} \Delta\alpha^{f,-}_{\ell} - \Delta\alpha_n^{\theta,-} + \sum_{\ell\in\mathcal{L}_n} B_\ell \mathcal{M}^{\pi}_n
  \leq 0. \label{eq:subdiff:angle2}
\end{align}
Hence, we have $\Delta\alpha_n^{+} \leq \mathcal{M}^\pi_n \leq \Delta\alpha_n^{-}$. Because $\Delta\bar\alpha_n^+$ and $\Delta\bar\alpha_n^-$ are the smallest incremental bid prices at node $n\in\bar{\mathcal{\mathcal{N}}}$, we obtain the bound $-\Delta\bar\alpha_n^+ \leq \mathcal{M}_n^{\pi} \leq \Delta\bar\alpha_n^-,\; n\in\bar{\mathcal{N}}$. \Halmos
\endproof

The price distortion is bounded for every node $\n\in\mathcal{N}$. Moreover, if the penalty parameters $\Delta\alpha_{\ell}^{f,+},\Delta\alpha_{\ell}^{f,-},\Delta\alpha_n^{\theta,+},\Delta\alpha_n^{\theta,-}$ are made arbitrarily small, then the price distortion at every node becomes arbitrarily small.  We now state results that are natural extensions of Theorems \ref{th:singlebound} and \ref{th:singlemedian}. 

\begin{theorem}\label{th:networkbounds}
Consider the stochastic clearing model \eqref{eq:stochangle}, and let the assumptions of Theorem \ref{th:networkdistortion} hold. The day-ahead quantities and phase angles are bounded by the real-time quantities and phase angles as
\begin{align*}
  \min_{\omega \in \Omega}D_j(\omega) \leq d_j\leq \max_{\omega \in \Omega}D_j(\omega),& \quad j\in\mathcal{D}\\
  \min_{\omega \in \Omega}G_i(\omega) \leq g_i\leq \max_{\omega \in \Omega}G_i(\omega),& \quad i\in\mathcal{G}\\
  \min_{\omega \in \Omega}F_\ell(\omega) \leq f_\ell \leq \max_{\omega \in \Omega}F_\ell(\omega),& \quad \ell\in\mathcal{L}\\
  \min_{\omega \in \Omega}\Theta_n(\omega) \leq \theta_n \leq \max_{\omega \in \Omega}\,\Theta_n(\omega),& \quad n\in\mathcal{N}.
\end{align*}
\end{theorem}

\proof{Proof}
For the suppliers and demands, we can use the same  procedure used in the proof of Theorem \ref{th:singlebound}. The bounds on the day-ahead flows and phase angles follow the same argument as well. We use the definition \eqref{eq:sumofsubdiffs} for simplicity. Consider the following two cases:
\begin{itemize}
  \item Case 1: The price distortion hits the lower bound for node $n$; we thus have $\mathcal{M}^{\pi}_n = -\Delta\alpha_n^{+}$. This implies that $-\sum_{\ell\in\mathcal{L}_n} \left( \Delta\alpha_\ell^{f,+} + \Delta\alpha_\ell^{f,-} \right) - \Delta\alpha_n^{\theta,+} - \Delta\alpha_n^{\theta,-} \in \mathcal{S}_n$ from \eqref{eq:subdiff:angle}, and hence we have
  \begin{subequations}
  \begin{align}
    & -1 \in \partial_{\theta_n}\mathbb{E}\left[\left( \Theta_n(\omega)-\theta_n \right)_+\right] \label{eq:angleub}\\
    & -1 \in \partial_{\theta_n}\mathbb{E}\left[\left(\sum_{n\in\mathcal{N}} B_{\ell n} (\Theta_n(\omega)-\theta_n) \right)_+\right], \quad \ell\in\mathcal{L}_n. \label{eq:flowub}
  \end{align}
  \end{subequations}
  From \eqref{eq:subdiff}, equation~\eqref{eq:angleub} implies that $\mathbb{P}\left(\Theta_n(\omega) \geq \theta_n\right) = 1$, and equation~\eqref{eq:flowub} implies that $\mathbb{P}(\sum_{n\in\mathcal{N}} B_{\ell n} \Theta_n(\omega) \geq \sum_{n\in\mathcal{N}} B_{\ell n} \theta_n) = \mathbb{P}(F_\ell(\omega)\geq f_\ell) = 1$ for $\ell\in\mathcal{L}_n$. Therefore, we have $\theta_n \leq \Theta_n(\omega),\;\omega\in\Omega$ and $\theta_n \leq \max_{\omega\in\Omega} \Theta_n(\omega)$. Similarly, $f_\ell \leq F_\ell(\omega),\;\forall\omega\in\Omega$ and $f_\ell \leq \max_{\omega\in\Omega} F_\ell(\omega)$ for $\ell\in\mathcal{L}_n$.
  \item Case 2: The price distortion hits the upper bound for node $n$; we thus have $\mathcal{M}^{\pi}_n = \Delta\alpha_n^{-}$. This implies $0 \in \mathcal{S}_n$ from \eqref{eq:subdiff:angle}, and hence we have
  \begin{subequations}
  \begin{align}
    & 0 \in \partial_{\theta_n}\mathbb{E}\left[\left( \Theta_n(\omega)-\theta_n \right)_+\right] \label{eq:anglelb}\\
    & 0 \in \partial_{\theta_n}\mathbb{E}\left[\left(\sum_{n\in\mathcal{N}} B_{\ell n} (\Theta_n(\omega)-\theta_n) \right)_+\right], \quad \ell\in\mathcal{L}_n. \label{eq:flowlb}
  \end{align}
  \end{subequations}
  From \eqref{eq:subdiff}, equation~\eqref{eq:anglelb} implies that $\mathbb{P}\left(\Theta_n(\omega) \leq \theta_n\right) = 1$, and equation~\eqref{eq:flowub} implies that $\mathbb{P}(\sum_{n\in\mathcal{N}} B_{\ell n} \Theta_n(\omega) \leq \sum_{n\in\mathcal{N}} B_{\ell n} \theta_n) = \mathbb{P}(F_\ell(\omega)\leq f_\ell) = 1$ for $\ell\in\mathcal{L}_n$. Therefore, we have $\theta_n \geq \Theta_n(\omega),\;\omega\in\Omega$ and $\theta_n \geq \min_{\omega\in\Omega} \Theta_n(\omega)$. Similarly, $f_\ell \geq F_\ell(\omega),\;\forall\omega\in\Omega$ and $f_\ell \geq \min_{\omega\in\Omega} F_\ell(\omega)$ for $\ell\in\mathcal{L}_n$.
\end{itemize}
\Halmos
\endproof

\begin{theorem}
Consider the stochastic clearing problem \eqref{eq:stochangle}, and let the assumptions of Theorem \ref{th:networkdistortion} hold. If the price distortions $\mathcal{M}_n^{\pi},\;n\in\mathcal{N}$ are zero at the solution, then
\begin{subequations}
\begin{align}
  d_j &= \mathbb{Q}_{D_j(\omega)}\left( \frac{\Delta\alpha_j^{d,-}}{\Delta\alpha_j^{d,+}+\Delta\alpha_j^{d,-}} \right),\; j\in\mathcal{D} \label{eq:convergencedj}\\ 
  g_i &= \mathbb{Q}_{G_i(\omega)}\left( \frac{\Delta\alpha_i^{g,+}}{\Delta\alpha_i^{g,+}+\Delta\alpha_i^{g,-}} \right), \; i \in\mathcal{G} \label{eq:convergencegi}.
\end{align}
\end{subequations}
\end{theorem}

\proof{Proof}
For \eqref{eq:convergencedj} and \eqref{eq:convergencegi}, we can use the same  procedure used in the proof of Theorem \ref{th:singlemedian}.\Halmos
\endproof

\begin{corollary}
If the incremental bid prices are symmetric from Corollary~\ref{thm:abscost}, then $d_j = \mathbb{M}\left(D_j(\omega)\right),\; j\in\mathcal{D}$, and $g_i = \mathbb{M}\left(G_i(\omega)\right),\; i\in\mathcal{G}$.
\end{corollary}

We treat the penalty terms purely as a means to constrain the day-ahead flows and phase angles and induce the desired pricing properties. Our results indicate that this can be done with no harm by allowing $\Delta\alpha^{f,+}_{\ell},\Delta\alpha^{f,-}_{\ell},\Delta\alpha^{\theta,+}_n,\Delta\alpha^{\theta,-}_n$ to be sufficiently small. Moreover, making these arbitrarily small guarantees that the expected social surplus of the stochastic problem \eqref{eq:surplussto} satisfies $\varphi^{sto}\approx \varphi$. The alternative is to simply impose day-ahead bounds of the forms \eqref{eq:det:flowbounds} and \eqref{eq:det:anglebounds} and to eliminate the penalty terms on the flows and phase angles. In this case, however, we cannot guarantee that the price distortions are bounded, as we illustrate in the next section. In addition, similar to the case of day-ahead quantities, imposing day-ahead bounds on flows would require us to choose a proper statistic for the bounds of flows and phase angles, which might not be trivial to do. 

We now prove revenue adequacy and zero uplift payments in expectation for the network-constrained formulation. We denote a minimizer of the partial Lagrange function \eqref{eq:lagstoch} (subject to the constraints \eqref{eq:boundstoch1}-\eqref{eq:boundstoch2}) as $d_j^*,D_j^*(\cdot),g_i^*,G_i^*(\cdot),\theta^*_n,\Theta_n^*(\cdot),\pi^*_n,\Pi^*_n(\cdot)$. Because the problem is convex, we know that the prices $\pi^*_n,\Pi^*_n(\cdot)$ satisfy
\begin{align*}
  (d_j^*,D_j(\cdot)^*,g_i^*,G_i^*(\cdot),\theta_n^*,\Theta^*_n(\cdot)) = \mathop{\textrm{argmin}}_{d_j,D_j(\cdot),g_i,G_i(\cdot),\theta_n,\Theta_n(\cdot)} \quad & \mathcal{L}(d_j,D_j(\cdot),g_i,G_i(\cdot),\theta_n,\Theta_n(\cdot),\pi^*_n,\Pi^*_n(\cdot))\nonumber\\
  \textrm{s.t.} \quad & \eqref{eq:boundstoch1}-\eqref{eq:boundstoch2}.
\end{align*}
Moreover, at $\pi^*_n,\Pi^*_n(\cdot)$, the partial Lagrange function can be separated as
\begin{align}\label{eq:lag2}
  &\mathcal{L}(d_j,D_j(\cdot),g_i,G_i(\cdot),\theta_n,\Theta_n(\cdot),\pi^*_n,\Pi^*_n(\cdot))=\nonumber\\
  & \qquad \sum_{i \in \mathcal{G}}\mathcal{L}_i^g(g_i,G_i(\cdot),\pi^*_n,\Pi^*_n(\cdot))+\sum_{j \in \mathcal{D}}\mathcal{L}_j^d(d_j,D_j(\cdot),\pi^*_n,\Pi^*_n(\cdot)) + \mathcal{L}^\theta(\theta_n,\Theta_n(\cdot),\pi^*_n,\Pi^*_n(\cdot)).
\end{align}
where the first two terms are defined in \eqref{eq:contlag} and
\begin{align}\label{eq:revlag}
  &\mathcal{L}^\theta(\theta_{\ell},\Theta_{\ell}(\cdot),\pi^*_n,\Pi^*_n(\cdot))=\nonumber\\
  &\sum_{\ell\in\mathcal{L}} \mathbb{E}\left[\Delta\alpha^{f,+}_{\ell} \left( \sum_{n\in\mathcal{N}} B_{\ell n} (\Theta_n(\omega)-\theta_n) \right)_+ + \Delta\alpha^{f,-}_{\ell} \left( \sum_{n\in\mathcal{N}} B_{\ell n} (\Theta_n(\omega)-\theta_n) \right)_- \right] \notag \\
  &+\sum_{n\in\mathcal{N}} \mathbb{E}\left[\Delta\alpha^{\theta,+}_n \left( \Theta_n(\omega)-\theta_n \right)_+ + \Delta\alpha^{\theta,-}_n \left( \Theta_n(\omega)-\theta_n \right)_- \right] \notag \\
  & -\sum_{n\in\mathcal{N}}\pi_n \left[ \sum_{\ell \in \mathcal{L}_n^{rec}} \left(B_{\ell n} \theta_n + B_{\ell,snd(\ell)} \theta_{snd(\ell)}\right) -\sum_{\ell \in \mathcal{L}_n^{snd}} \left(B_{\ell,rec(\ell)} \theta_{rec(\ell)} + B_{\ell n} \theta_n\right) \right] \notag\\
  & -\mathbb{E} \left[ \sum_{n\in\mathcal{N}} \Pi_n(\omega) \left(\sum_{\ell \in \mathcal{L}_n^{rec}} \left[ B_{\ell n} \left( \Theta_n(\omega) - \theta_n \right) + B_{\ell,snd(\ell)} \left( \Theta_{snd(\ell)}(\omega) - \theta_{snd(\ell)} \right) \right] \right.\right. \notag \\
  & \qquad \qquad \qquad \qquad \left.\left.-\sum_{\ell \in \mathcal{L}_n^{snd}} \left[ B_{\ell,rec(\ell)} \left( \Theta_{rec(\ell)}(\omega) - \theta_{rec(\ell)} \right) + B_{\ell n} \left( \Theta_n(\omega) - \theta_n \right) \right] \right)\right]
\end{align}
Consequently, one can minimize the partial Lagrange function by minimizing \eqref{eq:contlag} and \eqref{eq:revlag} independently.

\begin{theorem}\label{th:networkadequacy}
Consider the stochastic clearing problem \eqref{eq:stochangle}, and let the assumptions of Theorem \ref{th:networkdistortion} hold. Any minimizer $d_j^*,D_j(\cdot)^*,g_i^*,G_i(\cdot)^*,\theta^*_n,\Theta^*_n,\pi^*_n,\Pi^*_n(\cdot)$ of \eqref{eq:stochangle} yields zero uplift payments for all players and revenue adequacy in expectation:
\begin{subequations}
\begin{align}
  \mathcal{M}_i^U&=0, \quad i\in \mathcal{G}, \\
  \mathcal{M}_j^U&=0, \quad j\in \mathcal{D}, \\
  \mathcal{M}^{ISO}&\leq 0.
\end{align}
\end{subequations}
\end{theorem}

\proof{Proof}
For fixed $\pi^*_n,\Pi^*_n(\cdot)$, by the separation of the partial Lagrange function, the zero uplift payments directly result from Theorem~\ref{thm:zerouplift}. At fixed $\pi^*_n,\Pi^*_n(\cdot)$ we also note that $\theta_n=\Theta_n(\cdot)=0$ is a feasible candidate solution for the maximization of $\mathcal{L}^\theta(\theta_n,\Theta_n(\cdot),\pi^*_n,\Pi^*_n(\cdot))$ and that, at this suboptimal point, this term is also zero.

If the flow balances \eqref{eq:networkfwd} and \eqref{eq:networkrt} hold, we have
\begin{align}
  0 =
  & -\sum_{n\in\mathcal{N}}\pi_n \left[ \sum_{\ell \in \mathcal{L}_n^{rec}} \left(B_{\ell n} \theta_n + B_{\ell,snd(\ell)} \theta_{snd(\ell)}\right) -\sum_{\ell \in \mathcal{L}_n^{snd}} \left(B_{\ell,rec(\ell)} \theta_{rec(\ell)} + B_{\ell n} \theta_n\right) + \sum_{i \in \mathcal{G}_n} g_{i} - \sum_{j \in \mathcal{D}_n} d_j \right] \notag\\
  & -\mathbb{E} \left[ \sum_{n\in\mathcal{N}} \Pi_n(\omega) \left(\sum_{\ell \in \mathcal{L}_n^{rec}} \left[ B_{\ell n} \left( \Theta_n(\omega) - \theta_n \right) + B_{\ell,snd(\ell)} \left( \Theta_{snd(\ell)}(\omega) - \theta_{snd(\ell)} \right) \right] \right.\right. \notag \\
  & \qquad \qquad \qquad \qquad -\sum_{\ell \in \mathcal{L}_n^{snd}} \left[ B_{\ell,rec(\ell)} \left( \Theta_{rec(\ell)}(\omega) - \theta_{rec(\ell)} \right) + B_{\ell n} \left( \Theta_n(\omega) - \theta_n \right) \right] \notag \\
  & \qquad \qquad \qquad \qquad \left.\left. + \sum_{i \in \mathcal{G}_n} \left(G_{i}(\omega)-g_{i}\right) - \sum_{j \in \mathcal{D}_n} \left(D_{j}(\omega)-d_{j}\right) \right)\right].
\end{align}
Consequently, for any arbitrary set of prices $\pi_n,\Pi_n(\cdot)$, we have
\begin{align*}
  &-\sum_{n\in\mathcal{N}}\pi_n \left[ \sum_{\ell \in \mathcal{L}_n^{rec}} \left(B_{\ell n} \theta_n + B_{\ell,snd(\ell)} \theta_{snd(\ell)}\right) -\sum_{\ell \in \mathcal{L}_n^{snd}} \left(B_{\ell,rec(\ell)} \theta_{rec(\ell)} + B_{\ell n} \theta_n\right) \right] \notag\\
  & -\mathbb{E} \left[ \sum_{n\in\mathcal{N}} \Pi_n(\omega) \left(\sum_{\ell \in \mathcal{L}_n^{rec}} \left[ B_{\ell n} \left( \Theta_n(\omega) - \theta_n \right) + B_{\ell,snd(\ell)} \left( \Theta_{snd(\ell)}(\omega) - \theta_{snd(\ell)} \right) \right] \right.\right. \notag \\
  & \qquad \qquad \qquad \qquad \left.\left.-\sum_{\ell \in \mathcal{L}_n^{snd}} \left[ B_{\ell,rec(\ell)} \left( \Theta_{rec(\ell)}(\omega) - \theta_{rec(\ell)} \right) + B_{\ell n} \left( \Theta_n(\omega) - \theta_n \right) \right] \right)\right] \\
  &=\sum_{n\in\mathcal{N}}\pi_n\left(\sum_{i \in \mathcal{G}_n} g_{i} - \sum_{j \in \mathcal{D}_n} d_{j}\right)+\mathbb{E}\left[\sum_{n\in\mathcal{N}}\Pi_n(\omega)\left(\sum_{i \in \mathcal{G}_n} \left(G_{i}(\omega)-g_{i}\right) - \sum_{j \in \mathcal{D}_n} \left(D_{j}(\omega)-d_{j}\right)\right)\right] \\
  &=\sum_{i \in \mathcal{G}_n} \pi_{n(i)}g_{i} +\mathbb{E}\left[\sum_{i \in \mathcal{G}_n} \Pi_{n(i)}(\omega)\left(G_{i}(\omega)-g_{i}\right)\right] -\sum_{j \in \mathcal{D}_n} \pi_{n(j)}d_{j} - \mathbb{E}\left[\sum_{j \in \mathcal{D}_n} \Pi_{n(j)}(\omega)\left(D_{j}(\omega)-d_{j}\right)\right] \\
  &=\mathcal{M}^{ISO}.
\end{align*}

Therefore, we have
\begin{align*}
  0 \geq 
  &\mathcal{L}^\theta(\theta_n^*,\Theta_n^*(\cdot),\pi^*_n,\Pi^*_n(\cdot)) \\
  \geq 
  &-\sum_{n\in\mathcal{N}}\pi_n \left[ \sum_{\ell \in \mathcal{L}_n^{rec}} \left(B_{\ell n} \theta_n + B_{\ell,snd(\ell)} \theta_{snd(\ell)}\right) -\sum_{\ell \in \mathcal{L}_n^{snd}} \left(B_{\ell,rec(\ell)} \theta_{rec(\ell)} + B_{\ell n} \theta_n\right) \right] \notag\\
  & -\mathbb{E} \left[ \sum_{n\in\mathcal{N}} \Pi_n(\omega) \left(\sum_{\ell \in \mathcal{L}_n^{rec}} \left[ B_{\ell n} \left( \Theta_n(\omega) - \theta_n \right) + B_{\ell,snd(\ell)} \left( \Theta_{snd(\ell)}(\omega) - \theta_{snd(\ell)} \right) \right] \right.\right. \notag \\
  & \qquad \qquad \qquad \qquad \left.\left.-\sum_{\ell \in \mathcal{L}_n^{snd}} \left[ B_{\ell,rec(\ell)} \left( \Theta_{rec(\ell)}(\omega) - \theta_{rec(\ell)} \right) + B_{\ell n} \left( \Theta_n(\omega) - \theta_n \right) \right] \right)\right] \\
  &= \mathcal{M}^{ISO},
\end{align*}
where the second inequality holds because 
\begin{align*}
&\sum_{\ell\in\mathcal{L}} \mathbb{E} \left[\Delta\alpha^{f,+}_{\ell} \left( \sum_{n\in\mathcal{N}} B_{\ell n} (\Theta_n(\omega)-\theta_n) \right)_+ + \Delta\alpha^{f,-}_{\ell} \left( \sum_{n\in\mathcal{N}} B_{\ell n} (\Theta_n(\omega)-\theta_n) \right)_- \right]\\
&\qquad + \sum_{n\in\mathcal{N}} \mathbb{E}\left[\Delta\alpha^{\theta,+}_n \left( \Theta_n(\omega)-\theta_n \right)_+ + \Delta\alpha^{\theta,-}_n \left( \Theta_n(\omega)-\theta_n \right)_- \right]\geq 0.\Halmos
\end{align*}
\endproof

We highlight that the introduction of the penalty terms for flows does not affect revenue adequacy and cost recovery because the partial Lagrange function remains separable for fixed prices.



\section{Computational Studies}\label{sec:computation}

In this section, we illustrate the different properties of the stochastic model. We also demonstrate that the stochastic model outperforms the deterministic one in all the metrics proposed. We also seek to highlight stochastic formulations provide benefits that go beyond improvements in social surplus. The optimization problems considered in this section were solved using {\tt CPLEX-12.6.1}. All models can be accessed at \url{http://zavalab.engr.wisc.edu/data}. 

\subsection{System I}\label{sec:system1}

\begin{figure}
\FIGURE
{\includegraphics[width=4in]{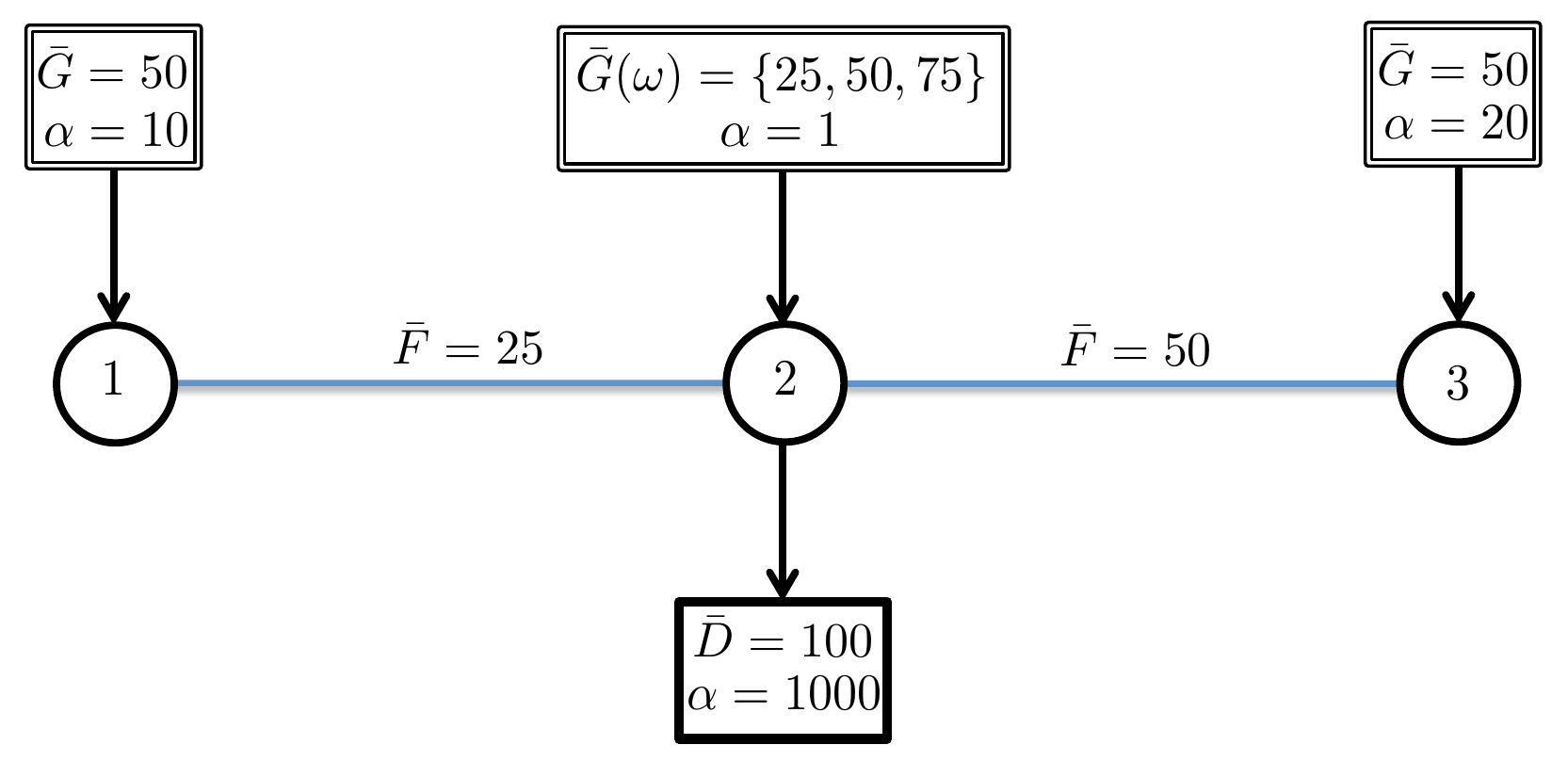}}
{Scheme of System I.\label{fig:systemI}}
{}
\end{figure}

We first consider System I sketched in Figure \ref{fig:systemI}. The system has two deterministic suppliers on nodes 1 and 3 and a stochastic supplier on node 2. The stochastic supplier has three possible capacity scenarios $G_2(\omega)=\{25,50,75\}$ MWh of equal probabilities $p(\omega)=\{1/3,1/3,1/3\}$.
For deterministic clearing, the day-ahead capacity limit $\bar{g}$ for the wind supplier will be set to 50 MWh, the expected value forecast. The demand in node 2 is deterministic at a level of 100 MWh. We use $\alpha^d=VOLL=$1000\$/MWh as the bid price and an incremental bid price of $\Delta \alpha^d=0.001$. The bid prices $\alpha^g_i$ for the suppliers are \{10,1,20\} \$/MWh, and the incremental bid prices $\Delta \alpha^g_i$ are \{1.0,0.1,2.0\} \$/MWh. The transmission capacities of lines $1\rightarrow 2$ and $2\rightarrow 3$ are deterministic and set to $\bar{F}_{1\rightarrow 2}=\{25,25,25\}$ MWh and $\bar{F}_{2\rightarrow 3}=\{50,50,50\}$, respectively, with the penalty of $\Delta\alpha^f = 0.001$. The line capacities have been designed such that the system becomes stressed in the scenario in which the stochastic supplier delivers only 25 MWh. We assume that the line susceptances are 50 for both lines. In this scenario, both transmission lines become congested, and real-time prices will reach high values. We use the penalty parameter $\Delta\alpha_n^{\theta} = 0.001$.

We compare the performance of the deterministic, stochastic here-and-now, and the stochastic wait-and-see (WS) settings. The results are presented in Table~\ref{tab:sysI_prices}. We compare the expected surplus for the suppliers $\varphi^{g}$ as well as prices and quantities. Because the demand is deterministic, the surplus for the consumers $\varphi^{load}$ is a constant. Consequently, we show only $\varphi^{g}$. For the deterministic setting, the expected supply surplus $\varphi^{g}$ is \$835, and the day-ahead prices $\pi_n$ are \{10,20,20\} \$/MWh. The price difference between the first two nodes results from the binding day-ahead flow for line $1\rightarrow 2$ line at 25 MWh. In the real-time market, the prices for each scenario $\Pi_n(\omega)$ are \{9,1000,22\}, \{9,20,20\}, and \{9,14,14\} \$/MWh with expected value $\mathbb{E}[\Pi_n(\omega)]=$\{9,345,19\}. There is a strong distortion in the prices, indicated by the metrics $\mathcal{M}^{\pi}_{avg}=109$ and $\mathcal{M}^{\pi}_{max}=325$.

\begin{table}
\TABLE
{System I. Comparison of quantities, prices, and social surplus.\label{tab:sysI_prices}}
{
\begin{tabular}{lcccccrr}
\toprule
  & $g_{i}$ & $\pi_n$ & $G_{i}(\omega)$ & $\Pi_n(\omega)$ & $\mathbb{E}[\Pi_n(\omega)]$ & $\varphi^{g}$ & $\mathcal{M}_{max}^\pi$ \\
\midrule
                &                &                & $\{25,25,50\}$ & $\{9,1000,22\}$ & & & \\
  Deterministic & $\{25,50,25\}$ & $\{10,20,20\}$ & $\{25,50,25\}$ & $\{9,20,20\}$   & $\{9,345,19\}$ & 835 & 325 \\
                &                &                & $\{25,75,0\}$  & $\{9,14,14\}$   & & & \\ 
\hline
                &                &                  & $\{25,25,50\}$ & $\{10,792,428\}$ & & & \\
  Stochastic    & $\{25,50,25\}$ & $\{10,276,154\}$ & $\{25,50,25\}$ & $\{10,20,20\}$   & $\{10,276,154\}$ & 835 & 0 \\
                &                &                  & $\{25,75,0\}$  & $\{10,15,15\}$   & & & \\ 
\hline
                & $\{25,25,50\}$ & $\{10,1000,929\}$ & $\{25,25,50\}$ & $\{10,1000,929\}$ & & & \\
  Stochastic-WS & $\{25,50,25\}$ & $\{10,20,20\}$    & $\{25,50,25\}$ & $\{10,20,20\}$    & $\{10,345,321\}$ & 800 & NA \\
                & $\{25,75,0\}$  & $\{10,14,14\}$    & $\{25,75,0\}$  & $\{10,14,14\}$    & & & \\
\bottomrule
\end{tabular}
}
{}
\end{table}

We now analyze the clearing of the stochastic formulation. The day-ahead prices are \{10,276,154\} and the real-time prices are \{10,792,428\}, \{10,20,20\}, \{10,15,15\} with expected value \{10,276,154\}. The price distortion metrics $\mathcal{M}^{\pi}_{avg}$, $\mathcal{M}^{\pi}_{max}$ are both zero. The stochastic WS (wait-and-see) solution is consistent in that it leads to no corrections of quantities in the real-time market and it yields the same day-ahead and real-time prices. Thus, {\em we can guarantee convergence of day-ahead and real-time prices for each scenario only in the presence of perfect information}. We note that the expected surplus as well as the day-ahead and real-time quantities for the stochastic and deterministic formulations are the same. The reason is that the {\em the deterministic and stochastic formulations have the same primal solution.} This situation might lead the practitioner to believe that no benefits are obtained from the stochastic formulation. The prices obtained, however, are completely different. Hence one can see that arguments based on social surplus do not fully capture the benefits of stochastic formulations.

The different prices obtained with both formulations lead to {\em drastically different payment distributions} among the market participants. As seen in Table \ref{tab:sysI_rev}, for the deterministic setting the suppliers obtain expected payments $\mathbb{E}[P^g_i(\omega)]$ of \$\{250,-7219,569\}. The wind supplier receives negative payments, and requires an uplift to enable cost recovery. In this case, the expected cost $\mathbb{E}[C^g_i(\omega)]$ for the wind supplier is \$52 and thus requires an expected uplift $\mathcal{M}_i^U$ of \$7,271. For the stochastic formulation, the expected payments are \$\{250,7321,7295\}. The wind supplier has positive payments and no uplift is required. This situation illustrates that the stochastic setting allocates resources efficiently. Note that all formulations are revenue adequate in expectation.

\begin{table}
\TABLE
{System I. Comparison of suppliers and ISO revenues.\label{tab:sysI_rev}}
{
\begin{tabular}{lccr}
\toprule
               & $\mathbb{E}[P^g_i(\omega)]$ & $\mathbb{E}[C^g_i(\omega)]$ & $\mathcal{M}^{ISO}$ \\
\midrule 
Deterministic  & \{250,{\bf -7219},569\} & \{250,52,533\} & -8400\\
Stochastic     & \{250,7321,7295\}       & \{250,52,533\} & -12719\\
Stochastic-WS  & \{250,9021,15656\}      & \{250,50,500\} & -9546 \\
\bottomrule
\end{tabular}
}
{}
\end{table}

\begin{remark}
The optimization problems for System I are highly degenerate, and thus multiple dual solutions (i.e., prices in our context) are available. We highlight that the solutions reported in this section were obtained from the barrier method (without crossover) implemented in {\tt CPLEX-12.6.1}, which would provide central point solutions. However, we also report the solutions obtained from different linear programming algorithms in Appendix.
\end{remark}

\subsubsection{Bounds on Day-ahead Quantities}

We now demonstrate that adding bounds on the day-ahead flows and phase angles, as opposed to adding penalty terms, can affect the pricing properties of the stochastic model. The price distortions $\pi-\mathbb{E}[\Pi(\omega)]$ obtained using day-ahead bounds (without the penalty term) are $\left\{-0.4,0,0\right\}$ while those obtained with the penalty term using a penalty of $\Delta \alpha^f=\Delta\alpha^\theta=0.001$ (without the bounds) are $\left\{0,0,0\right\}$. The penalty term achieves the desired pricing property. The day-ahead flows obtained with the penalty terms are $\{25,25\}$; these are the medians of the real-time flows which are $\{25,50\}$ for scenario 1, $\{25,25\}$ for scenario 2, and $\{25,0\}$ for scenario 3. This also implies that the day-ahead flows are bounded and therefore day-ahead bounds are redundant. Similarly, the day-ahead phase angles obtained with the penalty terms are $\{-3.58,-3.08,-3.58\}$; these are the medians of the real-time phase angles $\{-3.52,-3.02,-4.02\}, \{-3.58,-3.08,-3.58\}$, and $\{-3.65,-3.15,-3.15\}$ for scenarios 1, 2 and 3, respectively. This suggests that the day-ahead bounds bias the statistics.

\subsubsection{Effect of Incremental Bid Prices}

We experiment the effect that the incremental bid prices have on the price distortion. Consider the case in which the demand in the central node is also stochastic and with scenarios $D(\omega)=\{100,50,25\}$. We set the incremental bid price for the stochastic supplier $\Delta \alpha^g_2$ to $1.0$. For the demand incremental bid prices $\Delta \alpha^d = 0.001,0.01,0.1$, and 1.0, the maximum price distortions $\mathcal{M}_{max}^\pi$ are $0.001, 0.010, 0.069$, and 0.334, respectively. The distortion remains bounded by the incremental bid and can be made arbitrarily small as we decrease the incremental bid. The result is consistent with the properties established.

\subsection{System II}

\begin{figure}
\FIGURE
{\includegraphics[width=5in]{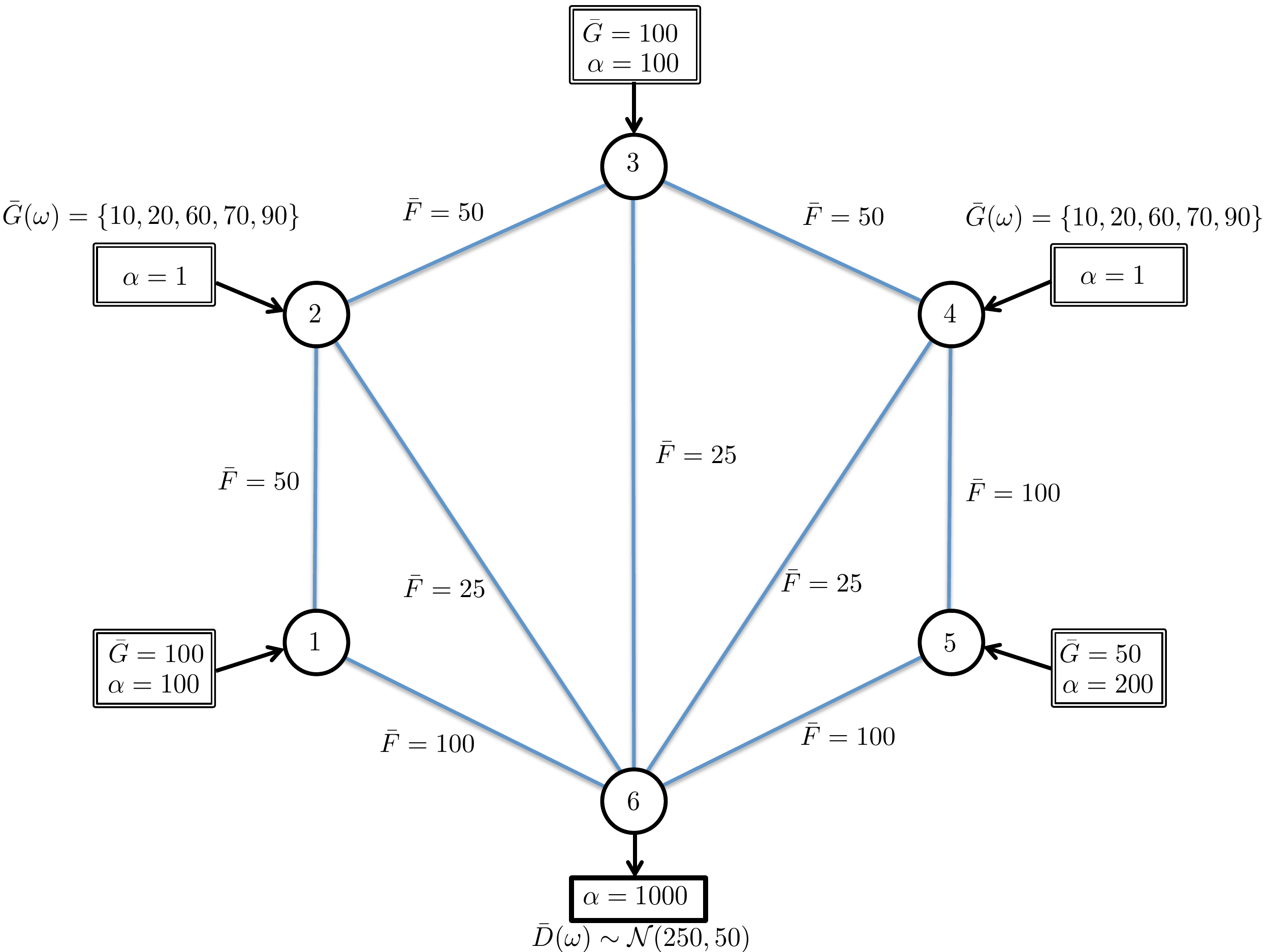}}
{Scheme of System II.\label{fig:systemII}}
{}
\end{figure}

We now consider the more complex system presented in Figure \ref{fig:systemII}. This is an adapted version of the system presented in \citet{philpott}. The system has two stochastic suppliers in nodes 2 and 4, three deterministic suppliers in nodes 1, 3, and 5 and one stochastic demand in node 6. The demand is treated as inelastic. The demand follows a normal distribution with mean 250 and standard deviation 50. We use sample average approximation for solving this model and generate 25 scenarios of the demand with equal probabilities. The stochastic suppliers can have 5 possible capacities $\{10,20,60,70,90\}$ MW. Each scenario represents one of the 25 different permutations from all possible capacities. The bid prices $\alpha_i^g$ for the suppliers are $\{100,1,100,1,200\}$ \$/MWh, and the incremental bid prices $\Delta\alpha_i^g$ are $\{10,0.1,10,0.1,20\}$. We set $\alpha^d=VOLL=1000$ \$/MWh and $\Delta \alpha^d=0.001$. We also set the penalty parameters $\Delta \alpha^f_\ell=\Delta\alpha_n^\theta=0.001$. We assume that all the line susceptances are 50.

\subsubsection{Price Distortion and Uplift Payments} 

\begin{table}
\TABLE
{System II. Comparison of day-ahead prices and surplus with deterministic and stochastic formulations.\label{tab:sysII}}
{
\begin{tabular}{lccc}
\toprule
                & $\varphi$ & $\mathcal{M}^{\pi}_n$      & $\pi_n$ \\
\midrule
  Deterministic & -139569   & $\{4,-28,-17,-158,-61,36\}$ & $\{100,-800,67,1,501,1000\}$ \\
  Stochastic    & -139569   & $\{-0.001,0,0,0,0,0\}$      & $\{100,-764,87,159,562,964\}$ \\
  Stochastic-WS & -139737   & NA & NA \\
\bottomrule
\end{tabular}
}
{}
\end{table}

The results are presented in Tables \ref{tab:sysII} and \ref{tab:sysII_rev}. We first note that the price distortion for the deterministic setting is large, reaching values as large as 158 \$/MWh. Also note that the distortion (premia) is small and positive in nodes 1 and 6 and large and negative in the other nodes. This is inefficient because it biases incentives towards a subset of players. The system is overly optimistic about performance in the real-time market where multiple scenarios exhibit transmission congestion, but the deterministic setting cannot foresee this. The stochastic formulation has almost the same expected social surplus as the deterministic formulation, but the price distortion is eliminated.

\begin{table}
\TABLE
{System II. Comparison of suppliers and ISO revenues with deterministic and stochastic formulations.\label{tab:sysII_rev}}
{
\begin{tabular}{ccccccc}
\toprule
                & $\mathbb{E}[P^{g}_i(\omega)]$    & $\mathbb{E}[C^{g}_i(\omega)]$      & $\mathcal{M}^{ISO}$\\ 
\midrule
  Deterministic & $\{8529,33,627,{\bf -1398},24927\}$ & $\{8529,0,627,20,9640\}$ & -129111\\
  Stochastic    & $\{8529,33,627,1903,27985\}$        & $\{8529,0,627,20,9640\}$ & -116885\\
  Stochastic-WS & $\{8458,37,570,1694,27714\}$        & $\{8458,0,570,20,9600\}$ & -117518\\
\bottomrule
\end{tabular}
}
{}
\end{table}

In Table \ref{tab:sysII_rev} we see that payments for both formulations are similar except for the 4th supplier, which is a stochastic supplier. This supplier receives a negative payment and requires uplift under deterministic clearing. The uplift is eliminated by using the stochastic formulation. The expected payments collected with the stochastic here-and-now solution are close to those of the perfect information solution.

\subsubsection{Convergence of Day-ahead Quantities: Quartiles vs. Means}

In Table~\ref{tab:sysII_quartile} we present the day-ahead quantities $g_i$ convergent to a quantile of the real-time quantities. The day-ahead quantities $g_i$ with the quartiles (i.e., quantiles at $p=0.25, 0.5, 0.75$) of the real-time quantities are compared with the means $\mathbb{E}[G_i(\omega)]$ of the real-time quantities. Recall that $\mathbb{Q}_{G_i(\omega)}(0.5) = \mathbb{M}[G_i(\omega)]$. We use asymmetric incremental bid prices $\Delta\alpha_i^{g,+}, \Delta\alpha_i^{g,-}$ as set in Table~\ref{tab:sysII_quartile}. As can be seen, convergence is achieved for all suppliers.    

\begin{table}
\TABLE
{System II. Day-ahead, quartiles of real-time quantities, and mean of real-time quantities for suppliers.\label{tab:sysII_quartile}}
{
\begin{tabular}{lllrrrrr}
\toprule
  $\Delta\alpha_i^{g,+}$ & $\Delta\alpha_i^{g,-}$ & Quantity & Gen 1 & Gen 2 & Gen 3 & Gen 4 & Gen 5 \\
\midrule
  $0.5 \Delta\alpha_i^g$ & $1.5\Delta\alpha_i^g$ 
    & $g_i$                            & 90   & 0   & 0   & 20   & 50 \\
  & & $\mathbb{Q}_{G_i(\omega)}(0.25)$ & 90   & 0   & 0   & 20   & 50 \\
  & & $\mathbb{E}[G_i(\omega)]$        & 84.6 & 0.4 & 5.7 & 19.7 & 48 \\
\hline
  $1.0\Delta\alpha_i^g$ & $1.0\Delta\alpha_i^g$ 
    & $g_i$                           & 91.7 & 0   & 0   & 20.8 & 50 \\
  & & $\mathbb{Q}_{G_i(\omega)}(0.5)$ & 91.7 & 0   & 0   & 20.8 & 50 \\
  & & $\mathbb{E}[G_i(\omega)]$       & 84.6 & 0.4 & 5.7 & 19.7 & 48 \\
\hline
  $1.5\Delta\alpha_i^g$ & $0.5\Delta\alpha_i^g$ 
    & $g_i$                            & 91.7 & 0   & 2.5 & 20.8 & 50 \\
  & & $\mathbb{Q}_{G_i(\omega)}(0.75)$ & 91.7 & 0   & 2.5 & 20.8 & 50 \\
  & & $\mathbb{E}[G_i(\omega)]$        & 84.6 & 0.4 & 5.7 & 19.7 & 48 \\
\bottomrule
\end{tabular}
}
{}
\end{table}

\subsubsection{Reliability Constraints}

We now consider the case in which there are random line failures. We consider 25 scenarios and assume that each one of the lines $1\rightarrow2$, $2\rightarrow6$, $3\rightarrow6$, $4\rightarrow6$, and $2\rightarrow3$ fails in at least five scenarios. All scenarios have equal probability. The results are presented in Table \ref{tab:sysII_rev2}. The deterministic setting becomes revenue inadequate in this case, whereas the stochastic setting is revenue adequate and achieves an expected ISO revenue that is close to that of the perfect information setting. An average price distortion of 1,374 \$/MWh  and a maximum distortion of 2,355 \$/MWh were obtained for the deterministic setting, indicating a pronounced effect of line failures on prices. In particular, we observed that several demands need to be curtailed in the deterministic case. The stochastic formulation eliminates the distortion and the need for uplift payments. Note also that the fourth wind supplier again faces a negative revenue under deterministic clearing and an uplift payment is needed. This again illustrates that deterministic clearing can affect resource diversification because it consistently biases the payments towards a subset of players.  

\begin{table}
\TABLE
{System II. Comparison of suppliers and ISO revenues with deterministic and stochastic formulations under transmission line failtures.\label{tab:sysII_rev2}}
{
\begin{tabular}{lccr}
\toprule
  & $\mathbb{E}[P^{g}_i(\omega)]$ & $\mathbb{E}[C^{g}_i(\omega)]$ & $\mathcal{M}^{ISO}$ \\
\midrule
  Deterministic & $\{88870,5,999,45196,63415\}$ & $\{2447,6,999,6,4240\}$ & {\bf 148311} \\
  Stochastic    & $\{1870 ,6,999,519  ,10296\}$ & $\{1870,6,999,5,3960\}$ & -39691 \\
  Stochastic-WS & $\{1700 ,5,908,516  ,10287\}$ & $\{1700,5,908,4,3600\}$ & -39965 \\
\bottomrule
\end{tabular}
}
{}
\end{table}

\subsection{IEEE-118 System}

We now demonstrate the properties of the stochastic setting in a more complex network. The IEEE-118 system comprises 118 nodes, 186 lines, 91 demand nodes, and 54 suppliers. We assume that three stochastic suppliers are located at buses 10, 65 and 112, and that each supplier has an installed capacity of 300 MWh. This represents 14\% of the total generation capacity. We also assume that a generation level for a given stochastic supplier follows a normal distribution with mean 300 MWh and standard deviation 150 MWh. The total generation capacity is 7,280 MW, and the total load capacity is 3,733 MW. We use sample average approximation and generate 25 scenarios for the stochastic suppliers. We use 10\% of the generation cost for the incremental bid prices $\Delta\alpha_i^g$. The demands are assumed to be deterministic, and we set $\Delta \alpha^d_j=0.001$. We use the penalty parameters $\Delta \alpha^f_{\ell}=\Delta\alpha_n^{\theta}=0.001$. 


\begin{table}
\TABLE
{IEEE-118 System. Comparison of suppliers and ISO revenues with deterministic and stochastic formulations.\label{tab:ieee118}}
{
\begin{tabular}{lrrrrr}
\toprule
  & $\mathcal{M}^U$ & $\varphi^{g}$ & $\mathcal{M}^{\pi}_{avg}$ & $\mathcal{M}^{\pi}_{max}$ & $\mathcal{M}^{ISO}$ \\
\midrule
  Deterministic & -151 & 36531 & 0.270 & 2.331 & -2306 \\
  Stochastic    & 0    & 36527 & 0.001 & 0.003 & -2009 \\
  Stochastic-WS & 0    & 36410 & 0     & 0     & -1982 \\
\bottomrule
\end{tabular}
}
{}
\end{table}

The results are presented in Table \ref{tab:ieee118}. The uplift payment and The price distortions exist for the deterministic setting. The stochastic formulation reduces the uplift payments by a factor of 4 and eliminates the price distortion ($\mathcal{M}^{\pi}_{max}=0.003$). The difference in social surplus between deterministic and stochastic formulations is marginal. Also note that the penalty parameters for the flows and phase angles can be set to arbitrarily small values because they have no economic interpretation. Consequently, they do not affect the social surplus significantly.

\section{Conclusions and Future Work}\label{sec:conclusions}

We have demonstrated that deterministic market clearing formulations introduce strong and arbitrary distortions between day-ahead and expected real-time prices that bias incentives and block diversification. We present a stochastic formulation capable of eliminating these issues. The formulation is based on a social surplus function that accounts for expected costs and penalizes deviations between day-ahead and real-time quantities. We show that the formulation yields day-ahead prices that are close to expected real-time prices. In addition, we show that day-ahead quantities converge to the quantile of real-time counterparts.

Future work requires extending the model in multiple directions. First, it is necessary to capture the progressive resolution of uncertainty by using multi-stage models and to incorporate ramping constraints and unit commitment decisions. Second, it is necessary to construct formulations that design day-ahead decisions that approach ideal wait-and-see behavior. \citet{morales2013pricing} demonstrate that this might be possible to do by using bi-level formulations, but a more detailed analysis is needed. Third, the proposed stochastic model is computationally more challenging than existing models available in the literature because it incorporates the detailed network in the first-stage. This leads to problems that much larger first-stage dimensions which are difficult to decompose and parallelize. Consequently, scalable strategies are needed.  Finally, it is necessary to explore implementation issues of stochastic markets such as effects of distributional errors.

\begin{APPENDIX}{System I Solutions from Different LP Algorithms}\label{app:system1}

Since System I problem is degenerate, multiple dual solutions are available. We report the solutions of System I obtained from different LP algorithms. Tables~\ref{tab:primalsimplex} and \ref{tab:dualsimplex} present the solutions from primal simplex method and dual simplex method, respectively, from using {\tt CPLEX-12.6.1}. Note that we present the dual solutions (i.e., prices) only, since all the primal solutions (i.e., day-ahead and real-time quantities) are identical. We also note that in all cases we have price converges. The deterministic settings result in the positive price distortions.

\begin{table}
\TABLE
{System I solutions from primal simplex method\label{tab:primalsimplex}}
{
\begin{tabular}{lccccrr}
\toprule
  & $\pi_n$ & $\Pi_n(\omega)$ & $\mathbb{E}[\Pi_n(\omega)]$ & $\mathbb{E}[P_i^g(\omega)]$ & $\mathcal{M}^{ISO}$ & $\mathcal{M}_{max}^\pi$ \\
\midrule
                &                & $\{9,1000,22\}$ & & & & \\
  Deterministic & $\{10,20,20\}$ & $\{9,22,22\}$   & $\{9,347,21\}$ & $\{250,-7183,533\}$ & -8400 & 327 \\
                &                & $\{9,18,18\}$   & & & & \\
  \hline
                &                & $\{10,22,22\}$ & & & & \\
  Stochastic    & $\{10,20,20\}$ & $\{11,20,20\}$ & $\{10,20,20\}$ & $\{250,967,533\}$ & -250 & 0\\
                &                & $\{9,18,18\}$  & & & & \\
  \hline
                & $\{10,1000,1000\}$ & $\{10,1000,1000\}$ & & & & \\
  Stochastic-WS & $\{10,20,20\}$     & $\{10,20,20\}$   & $\{10,347,347\}$ & $\{250,9167,16833\}$ & -8417 & NA \\
                & $\{10,20,20\}$     & $\{10,20,20\}$   & & & & \\
\bottomrule
\end{tabular}
}
{}
\end{table}

\begin{table}
\TABLE
{System I solutions from dual simplex method\label{tab:dualsimplex}}
{
\begin{tabular}{lccccrr}
\toprule
  & $\pi_n$ & $\Pi_n(\omega)$ & $\mathbb{E}[\Pi_n(\omega)]$ & $\mathbb{E}[P_i^g(\omega)]$ & $\mathcal{M}^{ISO}$ & $\mathcal{M}_{max}^\pi$ \\
\midrule
                &                & $\{9,1000,22\}$ & & & & \\
  Deterministic & $\{10,20,20\}$ & $\{9,20,20\}$   & $\{9,346,20\}$ & $\{250,-7183,533\}$ & -8400 & 326 \\
                &                & $\{9,18,18\}$   & & & & \\
  \hline
                &                & $\{10,22,22\}$ & & & & \\
  Stochastic    & $\{10,17,17\}$ & $\{10,20,20\}$ & $\{10,17,17\}$ & $\{250,767,533\}$ & -183 & 0\\
                &                & $\{11,10,10\}$ & & & & \\
  \hline
                & $\{10,1000,1000\}$ & $\{10,1000,1000\}$ & & & & \\
  Stochastic-WS & $\{10,20,20\}$     & $\{10,20,20\}$     & $\{10,343,343\}$ & $\{250,8917,16833\}$ & -8333 & NA \\
                & $\{10,10,10\}$     & $\{10,10,10\}$     & & & & \\
\bottomrule
\end{tabular}
}
{}
\end{table}

\end{APPENDIX}

\ACKNOWLEDGMENT{This work was supported by the U.S. Department of Energy, Office of Science, under Contract DE-AC02-06CH11357. We thank Mohammad Shahidehpour for providing the data for the IEEE-118 system. The first author thanks Paul Gribik, Tito Homem-de-Mello, and Bernardo Pagnoncelli for technical conversations and Antonio Conejo for providing references to existing work.}

\bibliographystyle{ormsv080}
\bibliography{markets}

\begin{flushright}
\scriptsize \framebox{\parbox{3.2in}{
The submitted manuscript has been created by UChicago Argonne, LLC,
Operator of Argonne National Laboratory ("Argonne").
Argonne, a U.S. Department of Energy Office of Science laboratory, is
operated under Contract No. DE-AC02-06CH11357. The U.S. Government retains
for itself, and others acting on its behalf, a paid-up,
nonexclusive, irrevocable worldwide license in said article to
reproduce, prepare derivative works, distribute copies to the
public, and perform publicly and display publicly, by or on behalf
of the Government.}} \normalsize
\end{flushright}

\end{document}